\documentstyle[12pt,aaspp4,flushrt]{article}

\begin{document}

\vspace{-0.7cm}

\title{The Disc Accretion in  Gravitational Field of a Rapidly
Rotating Neutron Star with a Rotationally Induced Quadrupol Mass Distribution }
\vspace{-0.1cm}
\centerline{The paper is accepted by {\it Astronomy Letters},v.24, p.894, 1998}
\author{N. R. Sibgatullin$^{1,3}$ and R. A. Sunyaev$^{2,3}$}

\newcommand{\beq}{\begin{equation}}
\newcommand{\eeq}{\end{equation}}

$1$ Moscow State University, Faculty of Mech. \& Math., Moscow, 119899 Russia

$^2$ Space Research Institute, Russian Academy of Sciences, ul. 
Profsoyuznaya 84/32, Moscow, 117810 Russia

$3$ Max-Planck Institut fuer Astrophysik, Garching, Germany

{\small
{\bf Abstract} We analyze the effect of the quadrupole component in the 
mass distribution of a rapidly rotating neutron star on energy release 
in the boundary layer on the surface of the accreting 
star and in the accretion disk in the cases where the stellar radius is 
smaller (or larger) than the radius of the marginally stable circular orbit.
We calculate the velocities and trajectories of the particles that fall 
on the stellar surface from the marginally stable orbit for a low-luminosity 
accreting source. The corresponding external gravitational field of the 
star is modeled by a new exact solution of the Einstein equations in 
vacuum. The parameters of this solution are adjusted by reconciling the 
numerical data for the radius of the marginally stable orbit and the 
gravitational redshift of Cook et al. (1994) with the corresponding 
data in the analytical solution. For various equations of state, we 
consider $1.4 M_{\odot}$ normal sequences and maximum mass normal sequences.}

\vspace{-0.7cm}

\section*{Introduction}

The discovery of quasi-periodic oscillations (QPOs) with a frequency 1 
kHz from several low-mass X-ray binaries and X-ray bursters has again 
attracted attention to the detailed theory of disk accretion onto 
neutron stars with weak magnetic fields. RXTE observations have clearly 
revealed accreting neutron stars with rotation periods of several milliseconds. There is no doubt that these objects emit through disk 
accretion, but so far no detailed theory has been developed that would 
fully explain their observed spectra. For a high luminosity $L > 0.05 
L_{edd}$, this problem requires a proper allowance for the inverse effect 
of radiation pressure on the trajectories of particles and on the shape 
of the disk and the boundary layer. In the low-luminosity limit $L << 
0.05 L_{edd}$, the inverse effect of radiation pressure can be ignored, 
and we can estimate the relative contributions of the disk and the 
boundary layer. Below, $L_{edd}$ denotes the critical Eddington stellar 
luminosity.

In the Newtonian theory of disk accretion onto a {\it nonrotating} star with 
no magnetic field, one half of the gravitational energy of accreted 
particles is liberated in an extended disk, and the other half converts 
into radiation and heat in a narrow equatorial layer around the star 
where the particle velocity decreases from the Keplerian velocity to the 
velocity of the star's outer layer.

For a {\it rotating} star in the Newtonian theory, the maximum energy that can 
be liberated in the boundary layer is proportional to the square of the 
difference between the angular velocities of particles in a Keplerian 
orbit with a radius close to the stellar radius and the stellar angular 
velocity $\Omega$:  $ \frac12 (\Omega_K-\Omega)^2 R^2\dot M =   \frac12
 G \dot M (1-\Omega/\Omega_K)^2/R$. , while the energy 
released in the disk does not depend on whether the neutron star rotates 
or not.

Granat, RXTE, and GRO observations of several X-ray bursters have shown 
that their spectra are highly variable and that hard tails occasionally 
appear in their spectra up to energies 100 keV. It is tempting to 
explain such hard tails by the emission from particles that break loose 
from the last stable orbit and fall on the stellar surface at a great 
velocity.

These two experimental facts (the discovery of QPOs and of hard X-ray 
emission from neutron stars) have prompted us to return to the problem 
that was stated by Sunyaev and Shakura (1986). These authors pointed out 
that the neutron-star radius could be appreciably {\it smaller} than the 
radius of the  marginally stable orbit. In this case, the accretion disk can 
extend only to the radius of the marginally stable orbit. The particles with a 
given energy and a given angular momentum they had in this orbit then 
fall on the stellar surface and release their energy through nuclear 

collisions and plasma effects in a layer with a surface density of less 
than $ 20 g/cm^2$.

Sunyaev and Shakura (1986) calculated the energy release in the disk and 
in the boundary layer on the stellar surface for a neutron star with 
negligible rotation. Kluzniak and Wagoner (1985) applied first-order 
corrections in rotation parameter for particle motion in a gap between 
the  marginally stable orbit and the stellar surface. In this approximation, 
Hartle and Thorne (1967) found the external gravitational field to be 
described by the Kerr linearized solution. Miller and Lamb (1996) made 
an attempt to model the radiation field of a bright thin annulus and 
estimated the effect of this radiation on particle motion.

The important phenomena, which was n't taken account in previous
works, is the oblateness of the fast rotating self
gravitating gaseous masses (like Macloren spheroids).
 
The gravitational field outside a rapidly rotating flattened neutron 
star differs markedly from the external field of black holes described 
by the Schwarzschild and Kerr solutions. For the gravitational field of 
a real star to be determined, the problem as a whole must be solved for 
a specified equation of state of the neutron star. It is quite clear 
that the complete formulation can be realized only by using powerful 
computational techniques and sophisticated numerical methods. The 
fullest known numerical results can be found in the fundamental paper by 
Cook et al. (1994). The tables in this paper give, in particular, the 
radii of the marginally stable orbit for corotating and counterrotating disks, 
equatorial radius, gravitational mass, angular momentum, forward and 
backward redshifts at the edges of the equator as a function of angular 
velocity of the neutron star for various equations of state of neutron
stars (A, AU, FPS, L, M) which were proposed, in particular, by 
Pandharipande (1971), Arnett and Bowers (1977), Friedman and 
Pandharipande (1981), Lorenz et al. (1993), and Wiringa et al. 
(1988). The stellar {\it rest} mass was fixed: for normal sequences, it was 
chosen so that the stellar gravitational mass was equal to $1.4 M_{\odot}$ in the 
static limit; for maximum mass normal sequences, it was chosen from the 
condition that stability be lost in the static limit. The results of 
Cook et al. (1994) were in agreement with the previous numerical 
results of Friedman et al. (1986).

In contrast to Sunyaev and Shakura (1986), Kluzniak and Wagoner (1985), 
and Miller and Lamb (1996), we use a global analytical approximation of 
the metric of the external field of a rapidly rotating neutron star by 
an exact solution of the Einstein equations in vacuum, which is entirely 
specified by the stellar mass $M$, its angular momentum $J$, and the 
quadrupole moment of the mass distribution $b$ (the multipole moments are 
defined in Sect. 3). For $b = 0$, this solution transforms to the Kerr 
solution. We determined the dependence of the quadrupole moment $b$ on 
the dimensionless rotation parameter $j =cJ/Gm^2$ for a given 
equation of state by comparing the analytically calculated radius of the 
marginally stable orbit with the numerical data of Cook et al. (1994). It 
turned out that there were essentially no multipole moments higher than 
the quadrupole in the external gravitational fields of rotating neutron 
stars.

The stellar rotation parameter cannot be greater than the critical value 
$j_{cr}$ at which the particle velocity on the stellar equator is equal to 
the Keplerian velocity at the same radius, because, in this case, 
centrifugal forces begin to strip matter off the star. On the other 
hand, Keplerian orbits are stable only if their radius is greater than 
the radius of the marginally stable orbit of corotating particles (this radius 
is less than or equal to $3r_g$). At critical values of the rotation 
parameter, the equatorial radius of a compact star is equal to the 
radius of the marginally stable Keplerian orbit. In this case, the region of 
free fall of accreted matter disappears. For $1.4 M_{\odot}$ normal sequences, 
the stellar equatorial radius can be smaller than the radius of the marginally
stable orbit only for relatively soft equations of state (EOS A, AU, 
FPS); the softer is the equation of state, the greater is $j_{cr}$ ($j_{cr} 
= 0.4$ for EOS A, $j_{cr} 
= 0.24$ for EOS FPS ).

In hundreds of millions (or maybe even billions) of years, accretion 
onto a neutron star can increase its mass up to a maximum possible value 
which becomes unstable in the static limit. For maximum mass normal sequences, 
the region of free fall of accreted matter exists for all equations of 
state; the critical rotation parameter is approximately equal to 0.65 
for the equations of state we considered (EOS A, AU, FPS, L). For 
maximum mass normal sequences, the external gravitational field differs 
only slightly from the Kerr field, while for  $1.4 M_{\odot}$ normal sequences, 
the deviation from the Kerr solution is significant.

The case of a corotating disk in binary systems is natural, because the 
accreted matter transfers its angular momentum to the star, forcing it 
to rotate in the same sense. However, the case of a counterrotating disk 
is also possible: if a binary system in a globular cluster undergoes a 
strong interaction (a collision) with a third star, it can knock one of 
the stars out of the binary and form a new pair with the remaining star
in which the rotation of the disk is opposite to that of the neutron 
star.

For particles that rotate in Keplerian orbits in the sense opposite to 
the stellar rotation (counterrotation), the radius of the marginally stable 
orbit (greater than $ 3r_g$) matches the stellar radius at $j=j_{cr-}$. We assign the minus sign to $j$ for a counterrotating disk. For 
maximum mass normal sequences $ |j_{cr-}| \approx j_{cr}$.

Thus, we make an attempt to estimate the effect of realistic 
gravitational fields outside rapidly rotating neutron stars on energy 
release (we derive explicit formulas for them) when the accreted matter 
falls on the stellar surface for some equations of state of neutron 
stars. To this end, we systematically use the exact solution of the 
Einstein equations which describes the gravitational field of a rapidly 
rotating object with a quadrupole mass distribution (see Sect. 3 and 
Appendix 1). Here, we restrict our analysis to $1.4 M_{\odot}$ normal sequences 
in the static limit and to maximum mass equilibrium sequences. Using the above solution, we study: 
(i) the luminosities of the accretion boundary layer on the stellar 
surface and the entire disk; (ii) the dependence of the radial and 
azimuthal velocity components of the particles at the stellar surface 
when they fall from the marginally stable Keplerian orbit; and (iii) the dependence of the radius of the marginally stable 
Keplerian orbit and of the gravitational redshift on dimensionless
rotation parameter (see Sect. 1 and 3). For comparison, we give simple 
relations for the above functions when modeling the stellar external 
field by the Kerr solution in Sect. 2. Of course, we recognize that 
radiation begins to affect appreciably the dynamics of accretion at a 
high luminosity. Below, we provide our model results for luminosities 
from the accretion boundary layer that are much lower than the Eddington 
luminosity.

\section{Formulation of the problem for an arbitrary external axisymmetric
field of a rotating neutron star}

Space-time is stationary outside a rotating neutron star with axial 
symmetry, and its metric in the system of coordinates constructed on the 
orbits of Killing's vectors takes the form (Weil and Papapetru):
\beq
ds^2 = -f(dt - \omega d\phi)^2 + f^{-1}[ \rho^2 d\phi^2 + e^{2\gamma}(d\rho^2 + dz^2)] 
\eeq
 Here $\rho, z$ are the Weyl coordinates, for which takes place
$g_{0\phi}^2 - g_{00}g_{\phi\phi}= \rho^2.$ Using the
functions $ f, \omega $  we can introduce the single complex valued
function-Ernst potential ${\bf E} \equiv f + i\psi$  where the
function  $\psi$ is defined by Einstein equation
$$
\nabla \cdot (\frac{f^{2}}{\rho^2} \nabla \omega) = 0 \quad \Rightarrow\frac{f^2}{\rho}\frac{\partial \omega}{\partial \rho} = \frac{\partial \psi}{\partial z};\quad\frac{f^2}{\rho}\frac{\partial \omega}{\partial z} = -\frac{\partial \psi}{\partial \rho};
$$
Here for convenience  the operators of divergeance and of gradient
$\nabla\cdot, \nabla $ in the cilindrical coordinates
of the 3D euclidean space  are defined.

  From Enstein's equations in vacuum, Ernst (1967) derived the equation
:
$$
Re  {\bf E}  \,\nabla^2 {\bf E}  -(\nabla  {\bf E})^2 = 0,
$$

for which there is an efficient method of solution using a linear 
integral singular equation with a Cauchy-type kernel (Sibgatullin 1991).

   The free particles move along geodesics, conserving their energy and
   angular momentum (due to the symmetry properties of the background
   geometry):
$$
\frac{dx^i}{ds} k_i = -E, \qquad \frac{dx^i}{ds} m_i = L
$$
 where the Killings vectors $k_i, m_i$  have the components $(-f, f\omega, 0, 0)$ and
$(f\omega, \rho^2/f-f\omega^2, 0,0)$ respectively; 
\beq
\frac{dt}{ds} = E(\frac{1}{f}-\frac{f\omega^2}{\rho^2}) + L\frac{f\omega}{\rho^2},\qquad\frac{d\phi}{ds} =­- E\frac{f\omega}{\rho^2} + L\frac{f}{\rho^2}.
\eeq
Let us assume that the field is symmetric about the equatorial plane 
(this hypothesis can be proved in the Newtonian formulation; see Lamb 
1947). For the radial component of the 4-velocity vector of a particle 
moving in the equatorial plane, we then obtain from (1.2)
\beq
(\frac{dr}{ds})^2\frac{e^{2\gamma}}{f} = \frac{E^2}{f} - \frac{f}{\rho^2}(L - \omega E)^2 -1 \equiv V(\rho).
\eeq
\subsection{The equation for the radius of the marginally stable
keplerian orbit}
 The circular trajectories in the equatorial plane can be found with
the  help of the conditions $ V(\rho) =d V/d\rho = 0$, from where
 one can find the energy and the angular momentum of a circularly
 rotating particle as a function of its radius:
\beq
E=\frac{\sqrt{f}}{\sqrt{1-f^2x^2/r}}, \quad L=E(p+\omega),
\eeq
$$p\equiv r(-l+\sqrt{l^2+m-m^2r})/n,\quad l\equiv f\dot{\omega},\,m\equiv\dot
{f}/f,\,n\equiv f-r\dot{f}.
$$
Here dote denotes the derivative with respect to $r\equiv \rho^2$.

  The radius of the
 marginally stable orbit can be found with the help of the additional
 condition  $\ddot{V}(r)=0$, or equivalently of the condition of a
 minimum of energy on the circular orbits. In the explicit form it
 is
\beq
r(lmn+\dot{l} n-\dot{n} l)\sqrt{l^2+m-m^2r}-(l^2+m-m^2)(n-r\dot
{n})-rn(l\dot{l}+0.5(\dot{m}-m^2)-rm\dot{m})=0
\eeq
\subsection{Energy release in the accretion disc and in the boundary layer} 
   For the determination of the energy, released at the surface of
   the neutron star due to striking of the particles falling from the
   accretion disc, let us assume the particles to be at rest with
   respect to fluid particles in the outer layer of the star after giving back its
   angular momentum. 

  4-velocity of the equatorial fluid particle  in the coordinate system (1) has
   the components $q(1, \Omega, 0, 0)$ with
   $q \equiv
   \sqrt{f}/\sqrt{f^2-2f^2\omega\Omega-(r-f^2\omega^2)\Omega^2}.$ 
  The function  $q$ is equal to the component $u^0$ of 4 -velocity.It
   should be emphasized that angular velocity $\Omega$ is constant
   everywhere in the star  for the 
   case of 
   its uniform rotation.

 The energy per unit mass of such fluid particle is equal
\beq
 E_0
  \equiv f(1-\omega\Omega)\frac{\sqrt{f}}{\sqrt{f^2-2f^2\omega\Omega-(r-f^2\omega^2)\Omega^2}}.
\eeq
Let us assume that the kinetic energy of the falling particles relative 
to the fluid particles on the stellar surface completely transforms into 
radiation. In the Newtonian approximation, in which particles fall on 
the stellar surface from the Keplerian orbit, the emitted energy is 
$1/2GM(1-\Omega/\Omega_K)^2/R$. Indeed, the laws of conservation of 
energy and angular momentum for a neutron star when an accreted particle 
with a small mass m falls from a circular orbit on the stellar surface 
are
$$ \delta (I\Omega)=mR^2\Omega_K;\qquad \delta(I\Omega^2/2)=m\Omega_K
^2 R^2/2-l_s,$$
where $l_s$ is emitted energie. The change of the moment of inertia
$I$
is equal $\delta I=kmR^2.$  The coefficient is $k = 1$, if the particle 
remains in the boundary equatorial layer due to centrifugal forces. A more 
plausible assumption is that the accreted matter spreads over the 
stellar surface, causing the coefficient $k$ to decrease. For example, $k 
= 2/3$ for the spread over the surface of a sphere and $k = 2/5$ for a 
uniform mixing of the accreted matter within a sphere. It thus follows 
from the law of conservation of energy that  $I\delta
\Omega=mR^2(\Omega_K-k\Omega)$. Therefore, the change in the kinetic
energy is equal $\delta (I\Omega^2/2)=mk\Omega^2
R^2/2+m\Omega R^2(\Omega_K-k\Omega) = m\Omega_K ^2 R^2/2 -l_s$ . Hence, the formula for the 
emitted energy when a particle falls from a Keplerian orbit to the 
surface is $l_s = m R^2(\Omega_K-\Omega)^2/2 -(1-k)\Omega^2/2$. In what follow we assume the coefficient $k$ to be equal to
1. 

In general relativity, if the stellar radius is greater than the radius 
of the marginally stable orbit, the corresponding formulas for the luminosity 
from the accretion boundary layer on the stellar surface and from the 
entire disk surface are:
\beq
L_s = \dot{M}c^2(F -1)E_0,\qquad L_d = \dot M c^2 ( 1-E ), 
\eeq 
 where $ mc^2 E_0 $ is the energy of the particle lying on the star
surface, given by formula (4)  and $F\equiv (1-\Omega(p+\omega))qE$.  Here, we use the fact that the scalar 
product $F$ of the 4-velocity of the particle in a Keplerian orbit with 
the components  $E((1+f^2\omega p/r)/f,fp/r,0,0)$ and the 4-velocity of the 
particle on the stellar surface has the meaning of energy of a particle 
in a Keplerian orbit in a tetrad that is fixed relative to the surface 
of a rotating star.

When the radius of the marginally stable circular orbit is greater than the 
stellar radius, the energy emitted by the falling particles from the 
boundary layer per unit time is given by
\beq 
L_s = \dot M c^2 (F_* -1) E_0,\qquad F_*\equiv (1-\Omega(p_*+\omega_*))qE_*.
\eeq
Here
$m c^2 E_*$ is the energy of the particle on the marginally stable
orbit. 
 
 The 
gravitational energy emitted from the entire disk (bounded by the last marginally
stable orbit in this case) is
\beq
L_d = \dot{M}c^2(1 - E_*)
\eeq

In general relativity, the rate of increase of the stellar mass is given 
by
$$
\frac{dM}{dt}=\dot M E-L_s/c^2\quad \mbox{if}\quad R>R_*$$
$$
\frac{dM}{dt}=\dot M E_*-L_s/c^2 \quad\mbox{if}\quad R<R_*$$

Here, $E$ has the meaning of energy of a particle in a Keplerian orbit on 
the stellar surface [see formula (4)]; formula (7) must be used for $L_s$ 
in the first relation. For $R < R_*$ (in the second relation), $E_*$ is 
the particle energy in the last stable orbit, and formula (8) must be 
used for $L_s$. Thus, when calculating the rate of increase of the mass, 
we must take into account the losses by radiation, which can reach half 
the rest energy of the accreted particles (see the end of Sect. 3 for a 
discussion of this problem).

The rate of increase of the kinetic moment in general relativity is 
given by
$$\frac{dI\Omega}{dt}=\dot M L$$

The length of the stellar equator divided by $ 2\pi$ is

\beq
R=\sqrt{g_{\phi\phi}}=\rho/E_0.
\eeq

For this reason, we use below this circumferential stellar radius. In 
addition, this quantity, which has an invariant geometrical meaning, is 
used in theoretical studies of equilibrium of rotating neutron stars.

We emphasize that, in stationary and axisymmetric pseudo-Riemann spaces, 
the notions of energy and angular momentum have an exact meaning, and 
that the written expressions for the luminosities are given for the 
radiative energy that goes to pseudo-Euclidean infinity (if we ignore 
absorption in the disk).

 \subsection{Physical components of the particle's 4-velocity falling in the gap}
If the stellar equatorial radius is smaller than the radius of the marginally 
stable orbit, then the particles in the gap fall on the stellar surface 
in helical trajectories, conserving their energy and angular momentum 
they had in the marginally stable orbit. The physical components of the 
4-velocity of the particles in a ZAMO (zero angular momentum observer)
tetrad that does not rotate relative to 
a distant observer are given by
\beq
V_{\phi}=\frac{f(p_*+\omega_*)}{\sqrt{r}(1+f^2\omega (p_*+\omega_*-\omega)/r)},\eeq
\beq
V_{r}=\sqrt{1-\omega^2f^2/r}\frac{\sqrt{1-f^2(p_*+\omega_*-\omega)^2/r-(1-f_*^2p_*^2/r_*)f/f_*}}
{1+f^2\omega(p_*+\omega_*-\omega)/r},
\eeq
\beq
\frac{1}{\sqrt{1-V^2}}=\sqrt{f_*/f}\frac{1+f^2\omega
  (p_*+\omega_*-\omega)/r}{\sqrt{(1-\omega^2f^2/r)(1-f_*^2p_*^2/r_*)}}
\eeq
  \subsection{Gravitational redshift}
Let $Z_b, Z_f$  be the backward-equatorial  and the forward-equatorial redshift. The straightforward calculations gives for the
magnitude $Z\equiv(1+Z_b)(1+Z_f)$ the expression through the metric
coefficient $f$:
\beq
Z= 1/f
\eeq
\subsection{Some remarks}
Thus, the problem reduces to the choice of an appropriate geometry 
outside the rotating neutron star. For a global analytical approximation 
of the results of the numerous numerical calculations in the fundamental 
paper of Cook et al. (1994), we treat these data as follows (see Fig. 
1 and Tables 1-3). The dimensionless rotation parameter $j \equiv 
cJ/(GM^2)$ (here, $c$ is the speed of light; $G$ is the gravitational 
constant; and $J$ and $M$ are the angular momentum and mass of the neutron 
star, respectively) is plotted along the horizontal axis. The 
 dimensionless circumferential radius of the marginally stable orbit is plotted along the 
vertical axis; we assign the minus sign to $j$ for orbits that rotate in 
the sense opposite to the disk rotation and the plus sign for orbits 
that rotate in the same sense.

The treatment of three different theoretical equations of state of 
neutron stars for normal sequences with a rest mass of $1.4 M_{\odot}$ (EOS A, EOS 
AU, and EOS FPS) yields very close curves. The solid line corresponds to 
the circumferential radii of the marginally stable orbits in the Kerr solution 
with various rotation parameters. For the points that correspond to the 
evolutionary sequence of maximum mass normal sequences (MMNS) of the neutron stars (they are marked by 
filled triangles), we see a marked deviation from the above series of 
points for the normal sequences. Remarkably, the points essentially lie 
on the curve of the Kerr solution for $j$ in the interval (--0.15, 0.15); 
the agreement with the Kerr curve is better for the points of MMNS  . For larger absolute values of $j$, the 
deviation becomes significant. The deviation from the Kerr geometry is 
attributable to the appearance of a quadrupole component in the rapidly 
rotating star which is related to its flattening due to rapid rotation.

\section{The disc accretion in the Kerr field}
 
 Here, we use (as the first step and for comparison) the Kerr solution to 
describe the external geometry. In this case, the above formulas 
simplify appreciably. In contrast to Kluzniak and Wagoner (1985), Miller 
and Lamb (1996), Ebisawa et al. (1991), and Biehle and Blandford 
(1993), we obtain an explicit parametric representation of the 
quantities in question for an arbitrary rotation parameter.

The formulas for the energy and momentum of particles (divided by their 
mass) in circular orbits in the Kerr field are well known (Bardeen et
al. 1972):

\beq
E = \frac{1 - 2x + j x\sqrt{x}}{\sqrt{1 - 3x + 2j x\sqrt x}},\quad
L = M\frac{-2j x +(1 + j^2 x^2)/\sqrt{x}}{\sqrt{1 - 3x + 2j x\sqrt
x}},\quad x\equiv M/\tilde \rho
\eeq
Thus defined variable $x$ is used only in this section. Here, we adopt a system of units in which $G = c = 1,\tilde \rho$ is the 
radial Boyer-Lindquist coordinate in the Kerr metric. The radius of the 
marginally stable orbit can be determined from the condition of minimum of $E$ 
as a function of $x$. We will not express the radial coordinate $x$ of the 
marginally stable orbit as a function of $j$ (see an explicit formula with 
cubic radicals in Novikov and Frolov 1986), because this would lead to 
cumbersome expressions. Instead, we will consider the coordinate $x$ as a 
parameter and express all the quantities of interest in terms of this 
parameter. The expressions for $j_*, E_*$ and $L_*$ corresponding to the marginally 
stable orbit are
\beq
\label{f2}
j_* = \frac{4\sqrt{x} - \sqrt{3 - 2x}}{3x},\quad E_* = \sqrt{1 - \frac23 x},\quad L_*/M - j E_* = \frac{1}{x\sqrt{3}}.
\eeq
$x$ changes in the interval (1/9, 1/6) for the opposite senses of 
rotation of the disk and the neutron star and in the interval (1/6, 1) 
for the same senses of rotation. The corresponding values of $j$ for the 
former interval are negative.

Note that, in previous studies of disk accretion, the Schwarzschild 
metric was used with and without the first term of the expansion in $j$ 
(Sunyaev and Shakura 1986; Kluzniak and Wagoner 1985; Ebisawa et al.
1991; and Biehle and Blandford 1993). Our formulas (8) in the Kerr 
metric hold in any order in $j$, because they are exact. As follows from 
(4), the stellar circumferential radius is given by
\beq
R/M = x^{-1}\sqrt{1 + j^2 x^2(1 + 2x)}
\eeq

 The radial coordinate $r$, which was introdused in (4), 
is related to $x$ in the Kerr solution by
$$
r=M^2(1-2x+j^2x^2)/x^2
$$

In formulas (7)-(9) for the energy release in the entire accretion disk 
and in the accretion boundary layer, for the metric coefficients in the=
equatorial plane and the function $p$, which is given by (4) and which 
appears in all formulas for the luminosities, we must substitute their 
expressions in the Kerr solution:
$$
f=1-2x,\quad
f\omega=-2Mjx,\quad p=\frac{1-2x+j^2x^2}{\sqrt{x}(1-2x)(1-2x+jx\sqrt{x})}.
$$
For the marginally stable orbit, we obtain the following relations for these 
quantities:
$$f_*=1-2x_,\quad f_*\omega_*=2M(-4\sqrt{x}+\sqrt{3-2x})/3\quad
p_*=\frac43(\frac{1}{\sqrt{x}}-\frac{1}{\sqrt{3-2x}}).
$$
 For physical components of velocity  in ZAMO tetrad of the particle  falling in the
gap ( preseving energie and momentum corresponding to the circular
keplerian motion on the marginally stable orbit) one can obtain the formulas:
\beq
V_\phi =
\frac{2}{\sqrt{27}}\frac{\sqrt{1+j^2x^2-
 2x}(1+2\sqrt{(3-2x_*)/x_*})x}{\sqrt{1-2x_*/3}(1+j^2x^2)-2x^3 j/(x_*\sqrt{3})} 
\eeq

\beq
V_r = \frac{\sqrt{1+j^2
x^2(1+2x)}\sqrt{2(x-x_*)^3/(3x_*^2)}}{\sqrt{1-2x_*/3}(1+j^2x^2)-2x^3 j/(x_*\sqrt{3})},
\eeq

\beq
1/\sqrt{1-V^2}= 
\frac{\sqrt{1-2x_*/3}(1+j^2x^2)-2x^3 j/(x_*\sqrt{3})}{\sqrt{1+j^2x^2-
2x}\sqrt{1+j^2 x^2(1+2x)}}.
\eeq
It should be emphasized that the functions $V_{\phi}, V_r, \epsilon$
depending on   $R$ are given parametrically by (9), (11-13).
 
Finally, let us determine the trajectories of the particles that fall 
from the marginally stable circular orbit of the accretion disk on the stellar 
surface. To this end, we use the formulas for the first integrals of 
motion by dividing the expressions for the $ \phi$ and $r$ components of 
the 4-velocity by each other (Bardeen et al. 1972; Novikov and Frolov 
1986). We then obtain (by substituting for $j, E_*$ and $L_*$ their 
expressions from (8))
\beq
\frac{d\phi}{dx}= \frac{2(x-x_*)
  +\frac{4}{3}(x_*-\sqrt{(3-2x_*)x_*})}{\sqrt{2(x-x_*)^3}(1+j^2
  x^2-2x)}.
\eeq
 Substituting $x = x_*+t^2$, expanding the right-had part of (11) in 
elementary fractions, and performing the integration, we can easily 
derive a slightly cumbersome expression for $\phi =\phi(x)$ in 
explicit form.

Figure 2 shows the trajectories of the falling particles in the 
coordinates $ X\equiv R(x)\cos(\phi(x))/M, \quad Y\equiv
  R(x)\sin(\phi(x))$ are used, where, instead of $R(x)$,  the expression 
( 9 ) should be substituted.
 The extreme inner points on these spirals lie
  on the ergosphere.

\section{Reconstructing the parameters of a finite-multipole exact
  solution Einstein equations using numerical results of Cook et al.1994 }

The external gravitational fields in a stationary axisymmetric case are 
uniquely specified by Ernst's potential $E$ on the symmetry axis 
(Sibgatullin 1991). Ernst's potential for the Schwarzschild and Kerr 
solutions in Weil's coordinates is

$${\bf E}=\frac{(z-M)}{(z+M)},\quad {\bf E}=\frac{(z+ia-M)}{(z+ia+M)}.
$$ 
It is natural to seek the finite multipole solution at the axes of
symmetrie in the form
\beq \label{E}
 {\bf E}=\frac{z^n-Mz^{n-1}+\sum_{j=1}^n a_j z^{n-j}}
 {z^n+Mz^{n-1}+\sum_{j=1}^n a_j z^{n-j}}.
\eeq
The corresponding solution is symmetric about the equatorial plane, if 
we additionally require that the coefficients with the even subscript 
$a_{2k}$ be real (they are determined by the mass distribution and 
correspond to the Newtonian multipole moments), and that the 
coefficients with the odd subscript $a_{2k-1}$ be purely imaginary (they 
are determined by the distribution of angular momentum in the star and 
have no analog in the Newtonian theory). For this definition of the 
multipole moments, the Kerr solution is purely a dipole solution, and 
its higher multipoles are zero. For other definitions of the multipole 
moments, the Kerr solution has a fairly complex multipole structure 
(Hansen 1974; Thorne 1980; Kundu 1981; Simon and Beig 1983). For 
uniformly rotating stars, the coefficients $a_{2k-1}$ are most likely equal 
to zero for $k > 1$, with $a_1 = a \ne 0$; this coefficient is equal 
to the ratio of the star's angular momentum to its mass. For the fields 
outside a rotating neutron star, we assume that $a_{2n} =j^{2n}(\alpha_0^n +\alpha_1^n j^2 +
\ldots)$, where $j 
=a/M$ is the rotation parameter, i.e., the multipole coefficients 
are of the order $O(j^{2n})$. Therefore, the contribution of the higher 
multipole moments $a_{2n}$ at $n > 2$ is relatively small, because the 
rotation parameter varies in the range $ |j| < 0.65$ for realistic 
equations of state of neutron stars.

Using the method of constructing Ernst's potential in the entire space 
from its value on the symmetry axis (see Appendix 1), we obtain the 
solution corresponding to (22)
\beq
 {\bf E} =\frac{\Delta_-}{\Delta_+},\quad \Delta_{\pm}=
\det (E_{jk}^{\pm}),
\eeq
Here we introduce the notations
$$
E_{jk}^{\pm}=\frac{R_k}{\xi_k - a_j} \pm 1,\, E_{j+n,k}^{\pm}=\frac{1}{\xi_k
-a_j^{*}},\, R_k\equiv\sqrt{r+(z-\xi_k)^2}, j=\overline{1,n},\,k=\overline{1,2n}.
$$
 The constants $\xi_k, \,k=\overline{1,2n} $ are the roots of the real polynomial of the degree $2n$:
$$
 (z^n+\sum_{j=1}^n a_j z^{n-j})(z^n+\sum_{j=1}^n a_j^{*} z^{n-j})-M^2 z^{n-2}=0
,$$
the constants $a_j, j=\overline{1,n}$ are the roots of the denominator of the 
${\bf E}$ on the axes of the symmetry :$$z^n+Mz^{n-1}+\sum_{k=1}^{n} a_{k} z^{n-k}=0$$.

The metric coefficient $\omega$ is given by formula:
\beq
f\omega=2 \mbox{Im}\frac{\Delta_1}{\Delta_+},\quad \Delta_1=\det(\Omega_{jk})
\eeq
Here
$$
\Omega_{jk}= \frac{R_k}{\xi_k - a_j}+\frac{1+R_k+\xi_k-z}{M},\quad
\Omega_{j+n,k}= \frac{1}{\xi_k - a_j^{*}},\quad
j=\overline{1,n},\,k=\overline{1,2n}
$$
It should be remembered  that {\sl Im A} denotes the imaginary part of
$A; f$ is equal to the real part of $ E$. 

Here, we consider the important case $n = 2$ [it is the special case
of an exact solution to the 
system of Einstein-Maxwell equations for a rotating charged mass with 
quadrupole and magnetic moments found
 by Manko et al. (1994)]. For the quantities on the symmetry 
axis, we then have
\beq
{\bf E}=\frac{z^2+(ia-M)z+Mb}{z^2+(ia+M)z+Mb},\quad a\equiv Mj.
\eeq
The coefficient $b$ can be interpretated as a true quadrupole. The metric
 coefficients $f,\omega$ in the equatorial plane  are given by
\beq \label{f_omega}
f=\frac{A-B}{A+B},\quad\omega/M=-\frac{2jC}{A-B},\quad A\equiv
 (r_++r_-)^2r_+r_- - b,\quad B\equiv (r_++r_-)(r_+r_- - b+r).
\eeq
Here 
$$
C\equiv (r_+ + r_-)(r_-r_+ + r)+b,\quad
2r_{\pm}\equiv \sqrt{4r+(\sqrt{1-j^2}\pm\sqrt{1-j^2-4b})^2},\quad
r\equiv \rho^2/M^2
$$

Let us now calculate the radius of the marginally stable orbit by using 
formula (26) for the metric coefficients in the equatorial plane. We 
assume that the rotationally induced higher multipole moments $a_{2n}$ are 
of the order $O(j^{2n})$. Therefore, we represent the quadrupole 
coefficient $b$ as $b =j^2 k$, where $ k =\alpha_0 + \alpha_1 j^2$. The constants 
$\alpha_0,\quad \alpha_1$ are to be determined for each fixed rest mass 
and naturally depend on the choice of a specific equation of state for 
neutron stars. In this case, it will suffice to use numerical data for 
the radii of the marginally stable orbits only at two values of $j$.

Remarkably, all the other points from the numerical calculations fall on 
the derived theoretical relation between the radius of the marginally stable 
orbit and the rotation parameter constructed by using equation (5).

Thus, substituting (26) into (5) and finding its root $R = R_*$, we 
can determine the constants $\alpha_0,\quad \alpha_1$   by using the 
numerical values of $R_*$ only at two values of $j$ from Cook et al. 
(1994). By considering  1.4 $M_{\odot}$ normal sequences for some equations of state , we can adjust the$\alpha_0,\quad \alpha_1$ 
coefficients to fit the realistic numerical 
data as a whole with $ 5\%$ accuracy. Note that curve (15) for $R_*(j)/M$, 
which was constructed from the Kerr solution, differs markedly from the
realistic curves for $|j| > 0.15$. Curiously, the realistic curves for 
$R_*(j)/M$ and the Kerr curve (15), nevertheless, have a tangency of the 
first order at $j = 0$. This serves as a good illustration of the 
remarkable observation of Hartle and Thorne (1967) that the external 
gravitational field of a slowly rotating star is described by the Kerr 
metric linearized in rotation parameter. For our purpose, we reduced the 
data of Cook et al. (1994) for each of the three equations of state 
(EOS A, EOS AU, EOS FPS) and found that the numerical data of Cook et 
al. (1994) can be closely fitted by choosing $b = 2.3 j^2+1.3 j^4$ for 
EOS A, $b = 2.9 j^2+1.1 j^4$ for EOS AU, and $b = 3.2 j ^2+1.6 j^4$ for 
EOS FPS [see Tables 1-3 and Fig. 3 with a theoretical dependence 
$R_*(j)/M$ for the quadrupole solution with $\alpha_0,\quad \alpha_1$
corresponding to EOS AU and to the numerical results of Cook et al. 
(1994). For comparison, the curve that we constructed from the Kerr 
solution is also shown in this figure.

A comparison of the theoretical and numerical gravitational redshifts 
$Z(j) = 1/f$ (13), where $f$ is the corresponding metric coefficient 
on the stellar equator, can serve as another independent test of the 
validity of describing the external field of a rapidly rotating neutron 
star by the proposed quadrupole solution. Here, we do not have free 
parameters at all, because $\alpha_0,\quad \alpha_1$ were already fixed 
to reconcile the numerical and theoretical data for the radius of the 
marginally stable orbit. Substituting the numerical values of the stellar 
equatorial radius and the corresponding values of $j$ from Cook et al.
(1994) into the function $f$ (26) for a given equation of state, we can 
obtain the function $Z =Z(j)$ for the corresponding normal sequence. 
It can be verified that the values of $Z$ for the quadrupole solution 
with the above dependences of the quadrupole moment $b(j)$ differ from 
the numerical values of $Z(j) =(1+Z_b)(1+Z_f)$ from Cook et al. 
(1994) by 0.001 (see Fig. 4 for EOS AU and Tables 1--3). For 
comparison, the curve that corresponds to the Kerr solution is also 
shown here.

Let us now consider the case of MMNS for EOS A, 
EOS FPS, and EOS L. The maximum possible mass in the static limit is 
1.6541 $M_{\odot}$ (the rest mass is 1.9182 $M_{\odot}$ )  for EOS A, 1.7995  $M_{\odot}$ (the rest mass 
is 2.1028 $M_{\odot}$)   for EOS FPS, and 2.7002  $M_{\odot}$ (the
rest mass is 3.229 $M_{\odot}$) for 
EOS L (Cook et al. 1994). The corresponding neutron stars can 
"survive" only by rotating: they collapse when losing the angular 
momentum. Remarkably, these sequences have very similar dependences 
the quadrupole coefficient on rotation parameter: $b = 0.53 j^2+2.45 j^4$
(EOS A), $b = 0.7 j^2+2.3 j^4$ (EOS FPS), and $b = 0.6 j^2+2 j^4$ (EOS L). At 
these values of the quadrupole coefficient, the difference between the
theoretical and numerical radii of the marginally stable orbit is small for 
the same rotation parameter (see Fig. 5 for EOS A). Note that the 
difference between the curves for the quadrupole and Kerr solutions is 
smaller than that  for the normal sequences with $M = 1.4 M_{\odot}$ 
mentioned above.

Thus, the dimensionless parameters of MMNS  with distinctly different rest masses are, 
nevertheless, very similar. The usefulness of the dimensional theory can 
be shown for an almost universal dependence of the dimensionless angular 
momentum (rotation parameter) $j = cJ/GM^2$ on dimensionless angular 
velocity $W = \Omega GM/c^3$. For all three equations of state of 
state of MMNS the curves $j = j(W)$ that were 
constructed by using the numerical data of Cook et al. (1994) are 
essentially coincident (Fig. 6). Note that the functions $j =j(W)$ 
depend appreciably on the choice of an equation of state for 1.4  $M_{\odot}$ 
normal sequences (Fig. 7). Below, we show that the dimensionless 
luminosity from the accretion boundary layer and the physical velocity 
components of the falling particles on the stellar surface in the ZAMO 
tetrad as a function of rotation parameter depend only slightly on the
choice of an equation of state and are described fairly accurately by 
the Kerr curves for MMNS.

\section{Using the analytical solution for a fast rotating star with
the quadrupole mass distribution to describe the disc accretion}

We can now use our results for a realistic description of the external 
gravitational fields of rotating neutron stars to study the accretion 
caracteristics (6) and (8)-(11), which are intimately related to the 
geometry of space-time near rotating neutron stars. Using the stellar 
dimensionless equatorial radius from Cook et al. (1994) (see Table 
1-3), we can obtain realistic estimates for the rate of energy release 
and, accordingly, for the luminosity and the physical velocity 
components of the falling particles in the accretion boundary layer on 
the stellar surface. Clearly, these parameters depend on the rotation 
parameter, rest mass, and selected equation of state of neutron stars, 
because the equatorial radius itself depends on these parameters.

\subsection{The velocities of the falling particles on the star surface}

Let us first consider the dependence of the radial and azimuthal 
physical velocity components in the ZAMO tetrad of the particles that fall freely to the 
star from the marginally stable orbit for $1.4 M_{\odot}$ normal sequences in the 
static limit on the neutron-star surface. Here, the difference between 
the calculations that are based on the realistic quadrupole and Kerr 
geometries is pronounced (see Fig. 8 for EOS A). The curve $V_r(j)/c$ 
which corresponds to the "softer" equation of state differs by a smaller 
amount from the Kerr curve than do the curves with the "harder" 
equations of state, for which this difference is significant. Clearly, 
it is the radial velocity component that is responsible for the particle 
penetration deep into the neutron-star atmospheres. However, the 
azimuthal velocity component of the particles relative to the stellar 
surface mainly contributes to the energy release in the accretion 
boundary layer for a counterrotating disk as well. Incidentally, the differenc 
between the calculated azimuthal velocity component on the surface in 
the Kerr geometry and the realistic quadrupole component is less 
impressive than that for the radial velocity component (see Fig. 9 for 
EOS A). Below, we give quantitative estimates for the maximum radial 
velocity on the stellar surface (for a counterrotating disk) for the 
realistic quadrupole geometry:

$$V_{rmax}=0.119 c\quad \mbox{at}\quad W=-0.0484 \quad(\Omega=-6.94\times 10^3 1/sec)\, \quad\mbox{for EOS A}$$
$$V_{rmax}=0.097 c\quad \mbox{at}\quad W=-0.0475\quad  (\Omega=-6.797\times 10^3 1/sec)
\quad\mbox{for  EOS AU}$$ 
$$V_{rmax}=0.073\, c\quad \mbox{at}\quad W=-0.0431\quad (\Omega=-6.175\times 10^3 1/sec)
\quad\mbox{for  EOS FPS}$$  
 In the same time in Kerr geometry the maximum values of the radial component
of particles on the star surface are following
$$V_{rmax}=0.094 c\quad \mbox{at}\quad W=-0.0427\quad (\Omega=-6.12\times 10^3 1/sec)
\quad\mbox{for  EOS A}$$
$$V_{rmax}=0.068 c\quad \mbox{at}\quad W=-0.039\quad  (\Omega=-5.6\times 10^3 1/sec)
\quad\mbox{for  EOS AU}$$ 
$$V_{rmax}=0.047 c\quad \mbox{at}\quad W=-0.035\quad (\Omega=-5.01\times 10^3 1/sec)
\quad\mbox{for EOS FPS}$$  

As was already noted above, the calculations for MMNS yield a small difference between the physical parameters in 
the Kerr geometry and their values in the realistic quadrupole geometry 
(this can also be seen from Fig. 10 for EOS A):

$$V_{rmax}=0.26 c\quad \mbox{at}\quad W=-0.0573\quad (\Omega=-6.95\times 10^3 1/sec)\quad \mbox{for  EOS A}$$
$$V_{rmax}=0.277 c\quad \mbox{at}\quad W=-0.0655\quad  (\Omega=-6.22\times 10^3 1/sec)
\quad\mbox{for  EOS FPS}$$ 
$$V_{rmax}=0.284 c\quad \mbox{at}\quad W=-0.0677\quad (\Omega=-5.2\times1 10^3 1/sec)
\quad\mbox{for  EOS} L$$  

Compare these data with data in the Kerr geometry
$$V_{rmax}=0.25 c\quad \mbox{at}\quad W=-0.0446\quad (\Omega=-5.4\times 10^3 1/sec)\quad \mbox{for  EOS A}$$
$$V_{rmax}=0.255 c\quad \mbox{at}\quad W=-0.052\quad  (\Omega=-4.97\times 10^3 1/sec)
\quad\mbox{for  EOS FPS}$$ 
$$V_{rmax}=0.263 c\quad \mbox{at}\quad W=-0.0604\quad  (\Omega=-4.65\times 10^3 1/sec)
\quad\mbox{for EOS L}$$ 

Figure 11 gives an idea of the dependence of the azimuthal velocity 
component of the falling particles on the stellar surface for a 
MMNS with EOS A.

\subsection  {Energy Release in the Boundary Layer on the Stellar Surface}

Such large velocities of the falling particles on the stellar surface 
entail a substantial energy release in the accretion boundary layer. 
Figure 12 shows a plot of the dimensionless luminosity of the accretion 
boundary layer on the stellar surface $L_s/Mc^2$ against the 
dimensionless angular velocity for EOS A for a 1.4 $M_{\odot}$ normal sequence, and Fig. 13 shows the dependence of the dimensionless 
luminosity for a MMNS EOS A. Here, the 
difference between the curves that were constructed in the Kerr geometry 
and in the realistic quadrupole geometry is small. The luminosity is 
equal to zero for the critical dimensionless angular velocity $W_{cr}$, 
because, in this case, the particle velocity on the stellar surface 
reaches the velocity in a Keplerian orbit. At the same time, the stellar 
radius reaches the size of the marginally stable Keplerian orbit. The critical 
dimensionless angular velocity is given by
\beq
W_{cr}=\frac{f^2_* p_*/r_*}{1+f_*\omega_*/r_*}
\eeq
In the Kerr metric, we can explicitly express the critical stellar 
angular velocity in terms of $x$0 [see formulas (16)]:
\beq
W_{cr}=\frac{3x\sqrt{x}}{3+4x-\sqrt{3x-2x^2}} 
\eeq

This result indicates that, if the radius of a neutron star in the 
static limit is smaller than the radius of the last stable orbit $3r_g$, 
then its angular velocity cannot exceed its critical value, which is 
given by the above formulas. Note that the dimensionless critical 
velocity is completely determined by the dimensionless critical stellar 
angular momentum $j_{cr}$; this dependence is universal in the Kerr 
geometry. In the realistic quadrupole geometry, this dependence depends 
on the choice of an equation of state. For angular velocities greater 
than the critical value, centrifugal forces begin to strip the outer 
layers near the equator off the stellar surface. In the formula for 
$W_{cr}$, the functions (26) at $r =r*$ 
 must be substituted for $f$ and $g$. From this 
formula, we can derive the universal dependence $W_{cr}(j_{cr})$ for a given 
dependence of the quadrupole moment on the rotation parameter. Since the
radius of the marginally stable orbit is greater than $3 r_g$ for a 
counterrotating disk, the stellar equatorial radius increases to the 
size of the marginally stable orbit at negative values of $j_{cr-}$ that exceed 
$j_{cr}$ in absolute value. Thus, if the radius of a neutron star in the 
static limit is smaller than $3r_g$, then its rotation parameter can vary 
only in the range $|j| < j_{cr}$.

It follows from our results that, on the surface of a rotating neutron 
star, up to $50\%$ of the energy flux from the accreted matter $\dot M
c^2$ can be 
liberated for 1.4 $M_{\odot}$ normal sequences and up to $80\%$ 
of the energy flux $\dot M
c^2$   can be liberated for 
MMNS of neutron stars. Here, the nature invented the most perfect 
mechanism for the conversion of total energy of the falling particles 
(including their rest energy which is given by the Einstein formula $E 
= mc^2$) into radiation.

For the critical angular velocities of a counterrotating disk, the 
luminosity from the accretion boundary layer is a factor of 15 greater 
than the luminosity of the entire disk for EOS A (see fig.14) and is a factor of 12 
greater for the harder EOS FPS in the case of 1.4 $M_{\odot}$ normal sequences. 
For MMNS, which becomes unstable in the static 
limit, the luminosity from the accretion boundary layer can exceed the 
luminosity of the entire accretion disk by a factor of 20! (see fig.15). Thus, the 
luminosity from the compact region conspicuously dominates over the 
luminosity from the extended region. As was already noted above, for all 
the equations of state we studied, the dependence of the luminosities on
the dimensionless angular velocity is the same for MMNS.

\section{Concluding remarks}

Studies of the properties of apparent manifestations of neutron stars in 
general relativity reveals unexpected properties which are lacking in 
the Newtonian theory. One of such properties is the existence of a 
region of free fall of the accreted matter from the marginally stable 
Keplerian orbit (Sunyaev and Shakura 1986).

Another interesting fact is the relatively simple structure of the 
gravitational fields outside rapidly rotating neutron stars with 
relativistic equations of state. As we showed here, the quadrupole 
solution (see Sect. 3 and Appendix 1) that describes the external fields 
of these stars provides an excellent fit to the numerical data for the 
last stable orbit and for the gravitational redshift in the fundamental 
paper of Cook et al. (1994), which in turn are in agreement with the
numerical studies of other authors.

Our results suggest that the Kerr solution is inapplicable for 
describing the external gravitational fields of rapidly rotating neutron 
stars with 1.4$ M_{\odot}$ in the static limit. Particularly large differences 
are obtained for the radial velocity components on the stellar surface 
that were calculated in the Kerr geometry and in the realistic 
quadrupole geometry.

We showed that the ratio of the luminosities from the accretion boundary 
layer and the entire disc in the case of counterrotation can reach 12 
for normal sequences and 20 for MMNS. These 
results can be used in interpreting the strong X-ray emission from very 
compact regions.

We derived theoretical relations between the critical values of 
dimensionless angular velocity and dimensionless angular momentum when 
the neutron star become unstable to centrifugal forces.

From the widths of redshifted lines and from the average redshift, we 
can infer the angular velocity of the neutron star and the gravitational 
redshift at the stellar equator, respectively. Therefore, we can judge 
the stellar equation of state from independent observations of these 
quantities (as we showed here).

\section*{Acknowledgments}

We wish to thank I. Sibgatullin for numerous calculations and Dr. M.
Gilfanov for help and support, Dr. V.Astakhov and K.O'Shea for help in
the English translation.

\section*{References}

Arnett W.D. and Bowers, R.L.// Astrophys. J., Suppl. Ser., 1977, vol. 
33, p. 415.

Bardeen J.M., Press, W.H., and Teukolsky, S.A. 
//Astrophys. J.,@ 1972, vol. 178, p. 347.

Biehle G. and Blandford R.D. //Astrophys. J., 1993, vol. 411, p. 302.

Cook G.B., Shapiro
S.L., Teukolsky
 S.A. //Astrophys. J., 1994, vol. 424, p. 823.

Ebisawa K., Mitsuda K., and Hanawa T.// Astrophys. J., 1991, 
vol. 367, p. 213.

Ernst F. //Phys. Rev., 1968, vol. 167, p. 
1175.

Friedman B. and Pandharipande V.R.// Nucl. Phys., 1981, vol. 
A361, p. 502.

Friedman J.F., Ibser J.R., Parker L.// Astrophys. 
J., 1986, vol. 304, p. 115.

Hansen R.O.// J. Math. Phys., 1974, 
vol. 15, p. 46.

Hartle J.B. and Thorne K.// Astrophys. J., 1967, 
vol. 153, p. 807.

Kluzniak W. and Wagoner, R.V.// Astrophys. J.,
1985, vol. 297, p. 548.

Kundu, P.// J. Math. Phys., 1981, vol. 22, p. 
1236.

Lamb G., Hydrodynamics, Cambridge: Cambridge Univ. Press 1993, 6 Ed.

Lorenz C.P., Ravenhall D.G., and Pethick, C.J. 
//Phys. Rev. Lett., 1993, vol. 70, p. 379.

Manko, V.S., Martin, J.,
Ruiz E., Sibgatullin N.R., Zaripov M.N.//Phys. Rev. D, 1994, vol. 49,
p. 5144.

Miller
M.C. and Lamb F.K.// Astrophys. J., 1996, vol. 470, p. 
1033.

Novikov I.D. and Frolov V.P., Physics 
of Black Holes, Dordrecht: Kluwer 19896.

Pandharipande V.R. //Nucl. 
Phys., 1971, vol. A174, p. 641.

Sibgatullin N.R., Oscillations and 
Waves in Strong Gravitational and Electromagnetic Fields, Berlin: 
pringer, 1991.

Simon W. and Beig R.// J. Math. Phys., 1983, vol. 
24, p. 1163.

Sunyaev R.A. and Shakura N.I. //Pis'ma Astron. Zh.,
1986, vol. 12, p. 286.

Thorne K.S.// Rev. Mod. Phys., 1980, vol. 52, 
no. 2, pt. 1, p. 3.

Wiringa R.B., Fiks V., and Fabroccini A.
//Phys. Rev., 1988, vol. 38, p. 1010.

\section*{Appendix 1. Construction of the finite multipole solution,
possessing the plane of symmetry}

The solution of the Ernst equation with prescribed analytical data on
an interval at the axis :${\bf E}(\rho=0,z)\equiv e(z)$ can be
reduced
to the linear integral equation  with the singular kernel of
Cauchy type (Sibgatullin 1991) :
$$
\int\limits_{-1}^1
\frac{\mu(\sigma)(e(\xi)+\tilde{e}(\eta))}{(\sigma-\tau)\sqrt{1-\sigma^2}}\,d\sigma=0;\quad
\xi\equiv z+i\sigma\rho,\quad \eta\equiv
z+i\tau\rho,\,\sigma,\tau \in [-1, 1].\eqno (A 1)$$
Here the integral should be understood in the sense of principal
value; $e(\xi),\tilde{e}(\eta)$ are the analytical extentions in the
complex plane by the Riemann procedure of the functions $e(z), e*(z)$
given at the real axes.

 For uniformity the solution of   (A 1) must fulfill the 
normalisation condition:
$$
\int\limits_{-1}^1\frac{\mu(\sigma)\,d\sigma}{\sqrt{1-\sigma^2}}=\pi
\eqno (A 2)$$
The desired function ${\bf E}(\rho, z)$ then is expressed in the form:
$$
{\bf E}(\rho,
z)=\int\limits_{-1}^1\frac{\mu(\sigma)e(\xi)\,d\sigma}{\sqrt{1-\sigma^2}}.
\eqno (A 3)$$
One can represent the rational function $e(z)$ (\ref{E}) of finite multipole
solution in the form of the partial fractions:
$$e(z)=1 +\sum\limits_{j=1}^n\frac{e_j}{z-a_j} . $$
The solution of integral equation (A 1) can be sought in the form:
$$
\mu(\sigma)=A_0+\sum\limits_{k=1}^{2n}\frac{A_k}{\xi-\xi_k}. \eqno (A 4)$$

Having substituted (A 4) in the integral equation (A 1) and equated to
zero the coefficients by independant partial fractions one obtains
$$
\sum\limits_{k=1}^{2n}\frac{A_k}{(\xi_k-a_j^*)R_k}=1,\quad
A_0-\sum\limits_{k=1}^{2n}\frac{A_k}{\xi_k-a_j}=0,\,j=\overline{1,n} \eqno
(A 5).
$$
From (A 2) one gets:
$$
A_0+\sum\limits_{k=1}^{2n}\frac{A_k}{R_k}=1,\quad R_k\equiv\sqrt{r+\xi_k^2} \eqno (A 6)$$

The equations (A 5-6) form a closed system of linear algebraic
equations of $2n+1$ order with respect to unknown
$A_0,A_1,\dots,A_{2n}$.
According  (A 3) the solution $
{\bf E}(\rho,z)$ can be represented as follows
$$
{\bf E}(\rho,z)=1+\sum\limits_{k=1}^{2n}\frac{A_k}{R_k}\sum\limits_{j=1}^n\frac{e_j}{\xi_k-a_j}
\eqno (A 7)
$$
Having substituted the solution of the system (A 5-6) in the (A 7) one
obtains the solution, which can be represented in the form of the ratio of
the determinants $\Delta_{\pm}$ (see formula (23 )).

The coefficient $\omega$ in the metric form (1) can be found as
follows:
$$
\omega f=\frac{2}{\pi} {\sl
Im}\int\limits_{-1}^1\frac{\xi\mu(\sigma)\,d\sigma}{\sqrt{1-\sigma^2}}\eqno
(A 8)$$
Having substituted $\mu(\xi)$ in (A 8) by (A 4) one gets
$$
\omega f=-2\mbox{Im}(\sum\limits_{k=1}^{2n}A_k(1+\frac{\xi_k-z}{R_k})
$$
This expression can be represented in the form (24).

\section*{Appendix 2. Tables}

In the tables, the angular velocity is in units of  $10^3
 s^{-1}$, the mass 
is in solar masses, the circumferential radius is in  $G M c^{-2}$, the 
indices $\pm $ correspond to corotation and counterrotation, and the 
subscripts $k$ and $q$ characterize the corresponding quantities in the 
Kerr geometry and in our solution with the quadrupole moment chosen for 
each equation of state. $Z = (1+Z_b)(1+Z_f)$ (where $Z_b$ and $Z_f$ 
are the backward and forward redshifts, respectively) has the meaning of 
equatorial redshift. For comparison, columns 4-6 give the 
circumferential radii of the marginally stable orbit for the numerical 
solution of Cook et al. (1994), the Kerr solution, and the quadrupole 
solution, respectively. The circumferential radii in column 7, which 
were taken from Cook et al. (1994), are required to calculate the 
luminosity from the accretion boundary layer, the velocity component of 
the falling particles on the stellar surface, and the gravitational 
redshift. Column 8 gives theoretical gravitational redshifts.


\begin{table}
\begin{center}

{\bf Table 1. EOS A}

\vspace{3mm}

\begin{tabular}{cccccccc}\hline\hline
{$\Omega$}&
{$M/M_{\odot}$}&
{$j$}&
{$R_{*}/M$}&
{$R_{*}^{k}$}&
{$R_{*}^{q}/M$}&
{$R/M$}&
{$Z^{q}$}\\ \hline

6.94   & 1.417 & 0.45488  & 5.1936 & 4.48    &5.20267  &5.16 &1.664  \\

6.136  & 1.412 & 0.38617  & 5.198  & 4.71    &5.21374  &5.   &1.691 \\
 
5.019  & 1.408 & 0.3039   & 5.27   & 4.99    &5.28351  &4.86 &1.718 \\

3.244  & 1.403 &0.188978  & 5.4714 & 5.39    &5.47891  &4.73 &1.744 \\

0.     & 1.4   &0.        & 6.     & 6.      &6.       &4.66 &1.758\\ 

-3.244 & 1.403 & -0.1889  &6.6966  & 6.57    &6.68539  &4.73 &1.744 \\

-5.019 & 1.408 & -0.3039  &7.173   & 6.97    &7.15572  &4.86 &1.718 \\

-6.136 & 1.412 & -0.38617 & 7.5327 & 7.22    &7.50714  &5.   &1.691 \\
 
-6.94  & 1.417 &- 0.45488 & 7.8451 &7.43     &7.833777 &5.16 &1.664 \\

-7.545 & 1.421 & -0.5151  & 8.1237 & 7.62    &8.09965  &5.33 &1.632 \\

-7.954 & 1.425 & -0.56325 & 8.3567 & 7.76    &8.33841  &5.52 &1.601 \\

-8.236 & 1.428 & -0.60223 & 8.5492 & 7.87    &8.52761  &5.71 &1.570 \\

-8.431 & 1.431 & -0.63346 & 8.7042 & 7.97    &8.67937  &5.94 &1.538 \\

-8.591 & 1.434 & -0.66328 & 8.8592 & 8.03    &8.84255  &6.51 &1.466 \\ \hline
\end{tabular}
\end{center}
\end{table}
	
\begin{table}
\begin{center}

{\bf Table 2. EOS AU}

\vspace{3mm}

\begin{tabular}{cccccccc}\hline\hline
{$\Omega$}&
{$M/M_{\odot}$}&
{$j$}&
{$R_{*}/M$}&
{$R_{*}^{k}$}&
{$R_{*}^{q}/M$}&
{$R/M$}&
{$Z^{q}$}\\ \hline

4.457  & 1.407 & 0.31113  & 5.3411  & 4.96    &5.34405  &5.24 &1.633 \\

3.     & 1.403 & 0.20257  & 5.4807  & 5.33    &5.48028  &5.13 &1.649 \\

0.     &1.4    &0.        &6        &6        &6        &5.05 &1.660\\

-3     & 1.403 &- 0.20257 & 6.7787  & 6.65    &6.76061  &5.13 &1.649 \\

-4,457 & 1.407 &- 0.31113 & 7.2535  & 6.95    &7.2307   &5.24 &1.633 \\

-5.632 & 1.412 &-0.40860  & 7.7077  & 7.25    &7.6751   &5.40 &1.609 \\

-6.285 & 1.416 & -0.46983 & 7.9992  & 7.44    &7.9720   &5.54 &1.590 \\

-6.797 & 1.42  & -0.52340 & 8.2569  & 7.6     &8.2397   &5.69 &1.571 \\

-7.201 & 1.423 & -0.57129 & 8.4961  & 7.79    &8.4792   &5.86 &1.546 \\
  
-7.482 & 1.426 & -0.60883 &8.6835   & 7.84    &8.6668   &6.04 &1.524 \\

-7.713 & 1.429 & -0.64363 & 8.8572  & 7.98    &8.8594   &6.28 &1.495 \\
 
-7.9   & 1.432 & -0.67584 & 9.0285  & 8.08    &9.0293   &6.85 &1.432\\ \hline
\end{tabular}
\end{center}
\end{table}

\begin{table}[t]
\begin{center}

\vspace{10mm}

{\bf Table 3. EOS FPS}

\vspace{3mm}

\begin{tabular}{cccccccc}\hline\hline
{$\Omega$}&
{$M/M_{\odot}$}&
{$j$}&
{$R_{*}/M$}&
{$R_{*}^{k}$}&
{$R_{*}^{q}/M$}&
{$R/M$}&
{$Z^{q}$}\\ \hline

 2.284 &1.403    &0.20337  & 5.4932 & 5.33    &5.4969  &5.37 &1.602 \\
          
0.     &1.4      &0.       &6.      &6.       &6.      &5.27 &1.615\\

-2.284 & 1.403   & -0.2033 & 6.7927 & 6.65    &6.7753  &5.37 &1.602 \\

-4.112 & 1.406   & -0.2995 & 7.2220 & 6.93    &7.203   &5.49 &1.585 \\

-5.034 & 1.41    &-0.3807  & 7.6116 & 7.18    &7.5881  &5.64 &1.566 \\

-5.712 & 1.414   & -0.449  & 7.9508 & 7.4     &7.9276  &5.82 &1.544 \\

-6.175 & 1.417   & -0.5022 & 8.2241 & 7.57    &8.2023  &5.99 &1.523 \\

-6.544 & 1.421   & -0.5511 & 8.4795 & 7.67    &8.4615  &6.20 &1.500 \\

-6.842 & 1.424   & -0.5972 & 8.73   & 7.85    &8.7131  &6.45 &1.472 \\

-7.017 & 1.426   & -0.6287 & 8.8995 & 7.93    &8.8887  &6.70 &1.445 \\

-7.165 & 1.429   &-0.6593  & 9.0684 & 8.02    &9.0624  &7.36 &1.388 \\
 \hline
\end{tabular}
\end{center}
\end{table}

\newpage
\section*{Figure captions}
\indent

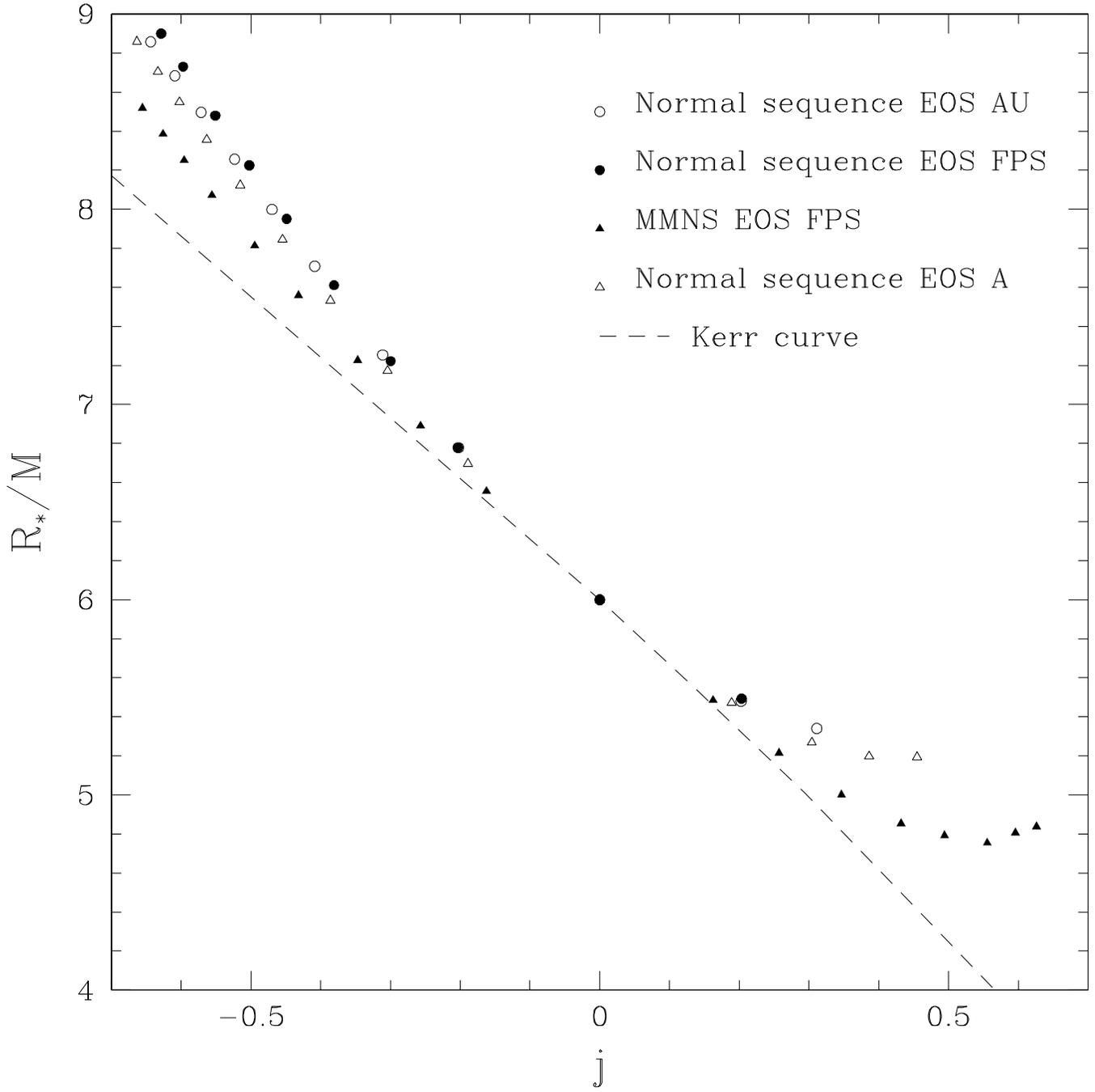
Fig. 1. The circumferential radius of the marginally stable orbit versus the 
dimensionless angular momentum. The solid line was constructed for the 
Kerr solution. The filled circles, open circles, and open triangles 
correspond to the numerical radii for 1.4 $M_{\odot}$ normal sequences in the 
static limit for EOS A, EOS AU, and EOS FPS. The filled triangles 
represent MMNS with EOS FPS.

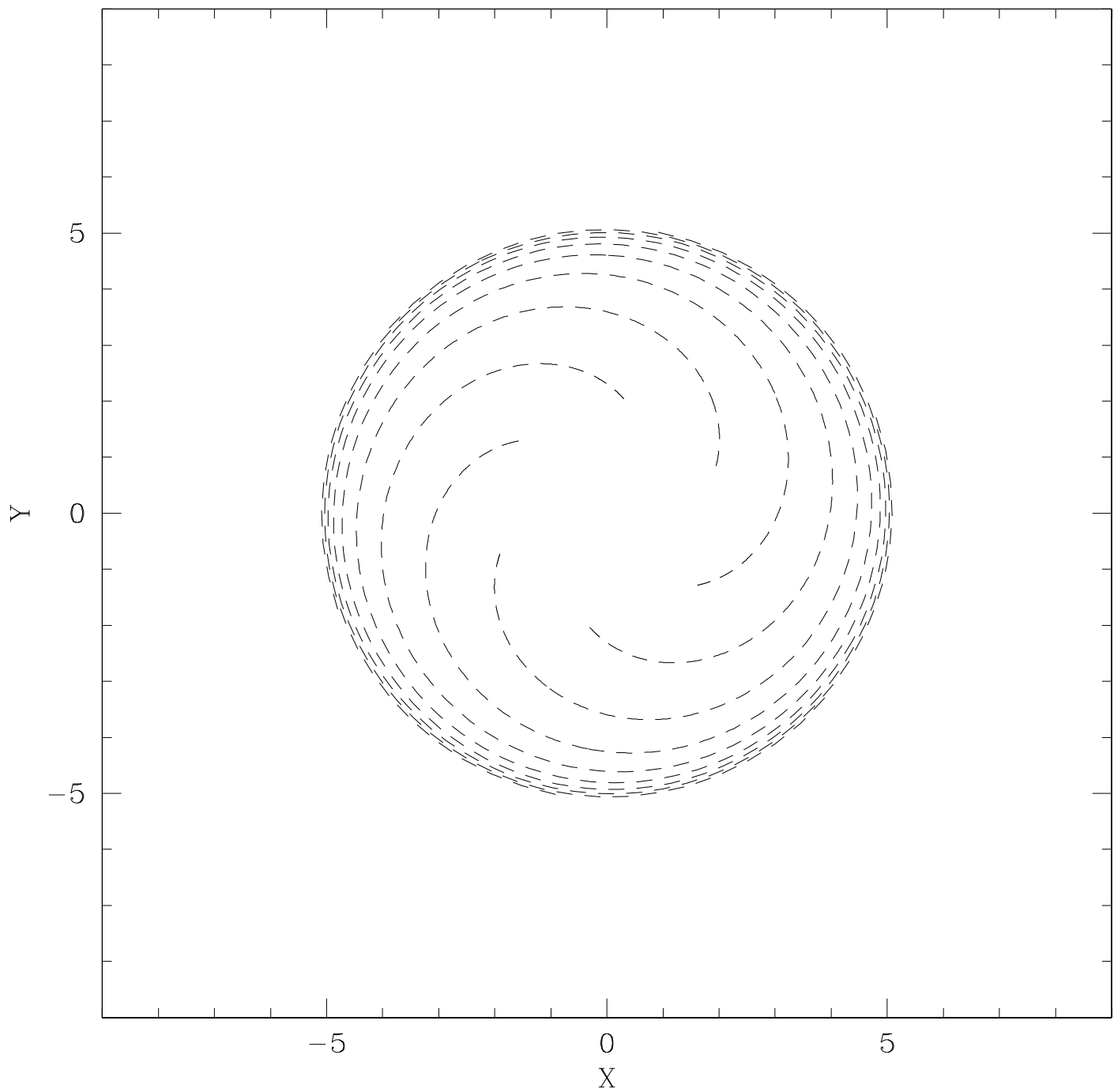
Fig. 2. The trajectories of the falling particles in the equatorial 
plane for a corotating disk for the rotation parameter $j = 0.22$. The 
innermost points correspond to the ergosphere.

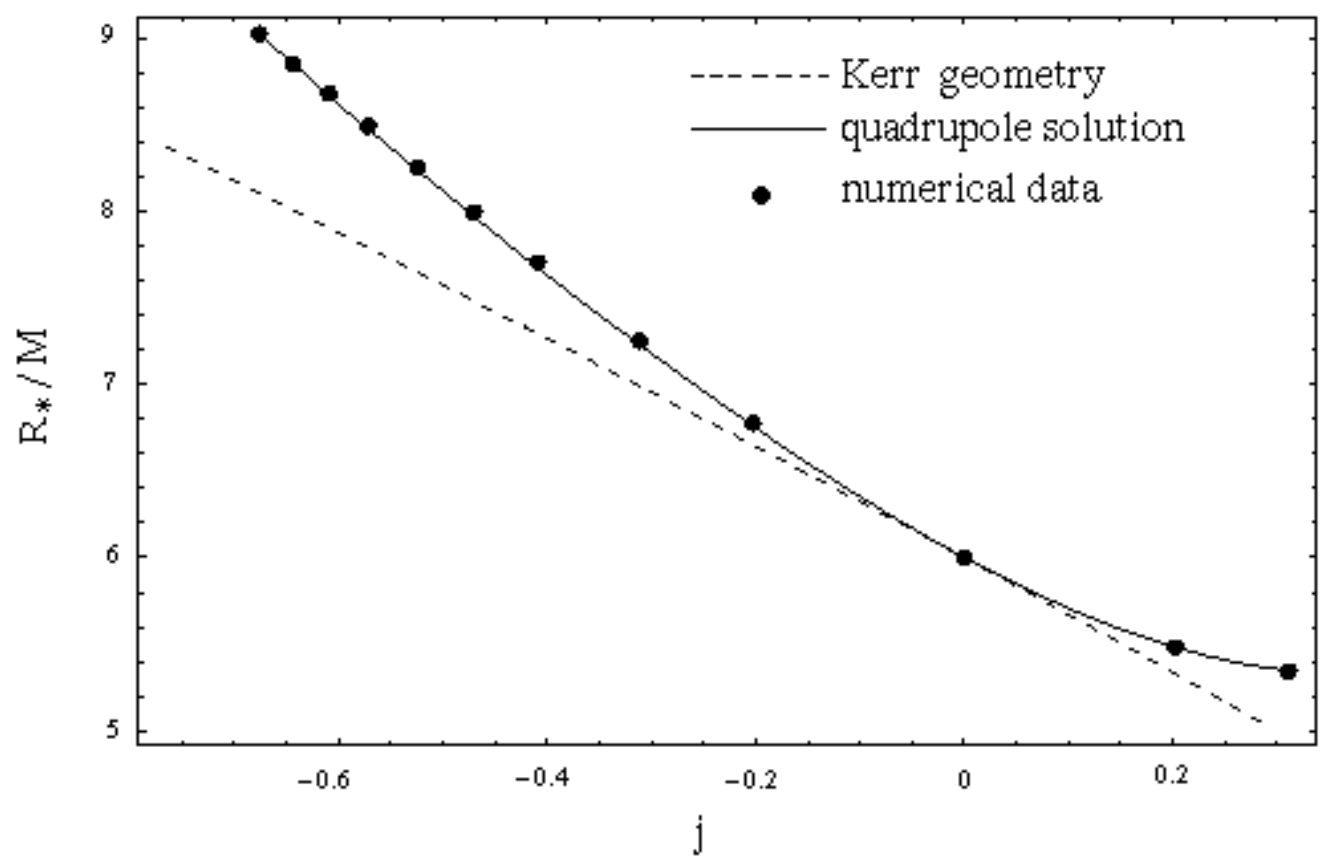
Fig. 3. The circumferential radius of the  marginally stable orbit versus the 
rotation parameter for a  1.4 $M_{\odot}$ normal sequence in the static limit and 
with EOS AU. Solid and dotted curves  correspond to the realistic quadrupole 
geometry and the Kerr solution, respectively; the dots represent the 
numerical data of Cook et al. (1994).

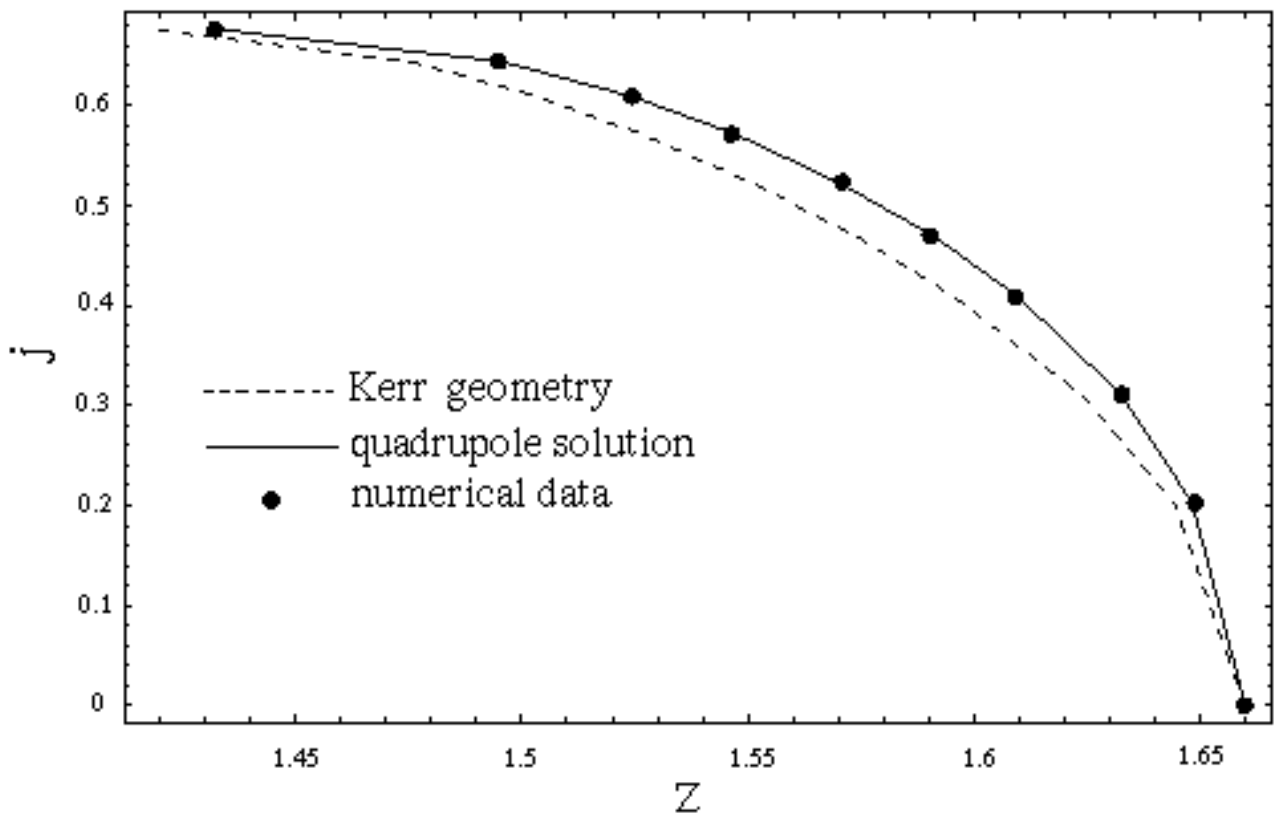
Fig. 4. The rotation parameter versus the gravitational redshift $Z =
(1+Z_b)(1+Z_f)$ for a  1.4 $M_{\odot}$ normal sequence in the static limit for 
EOS AU (solid curve ); dotted curve  corresponds to the Kerr solution; the dots represent the 
numerical data of Cook et al. (1994).

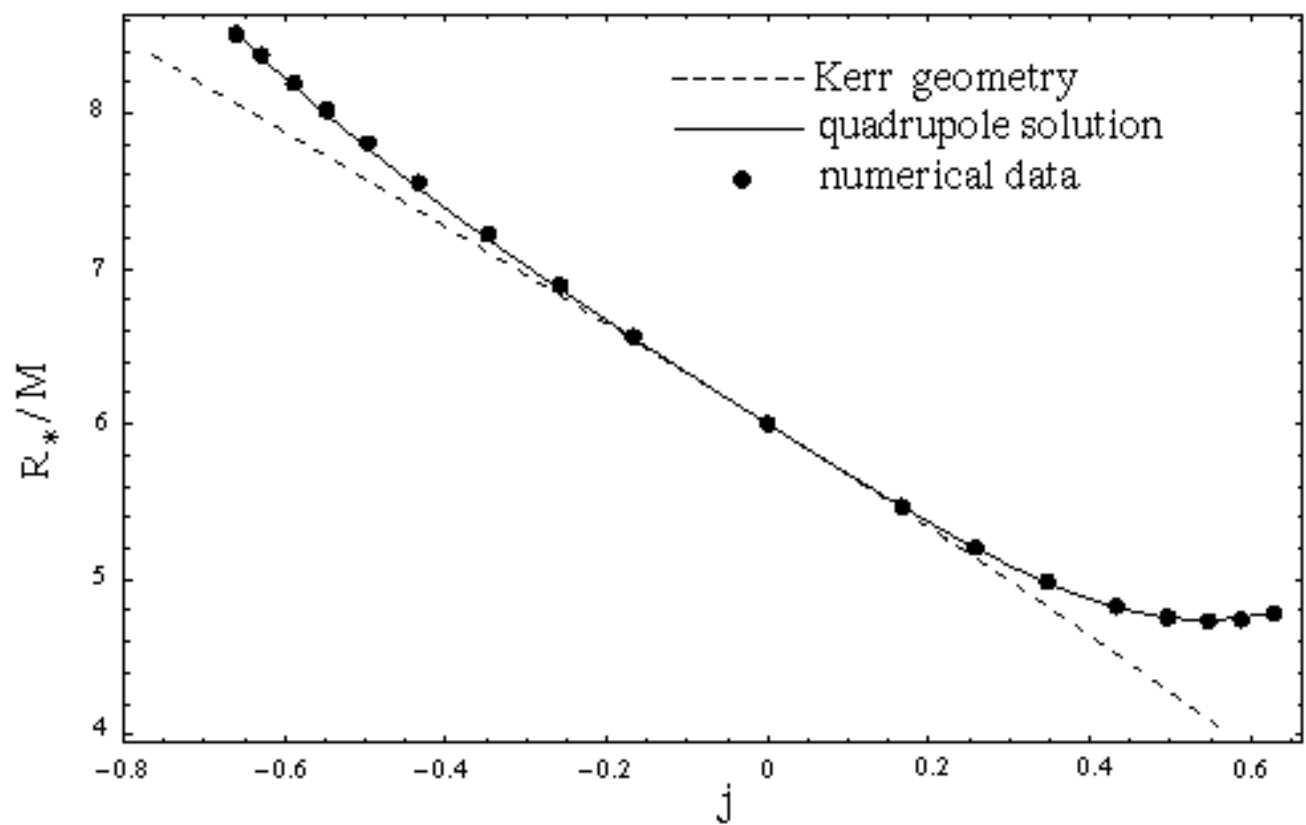
Fig. 5. The circumferential radius of the marginally stable orbit versus the
rotation parameter for a MMNS with 1.6551 $M_{\odot}$ in 
the static limit and with EOS A. Solid and dotted curves  correspond to the 
realistic quadrupole geometry and the Kerr solution, respectively; the dots represent the 
numerical data of Cook et al. (1994) .

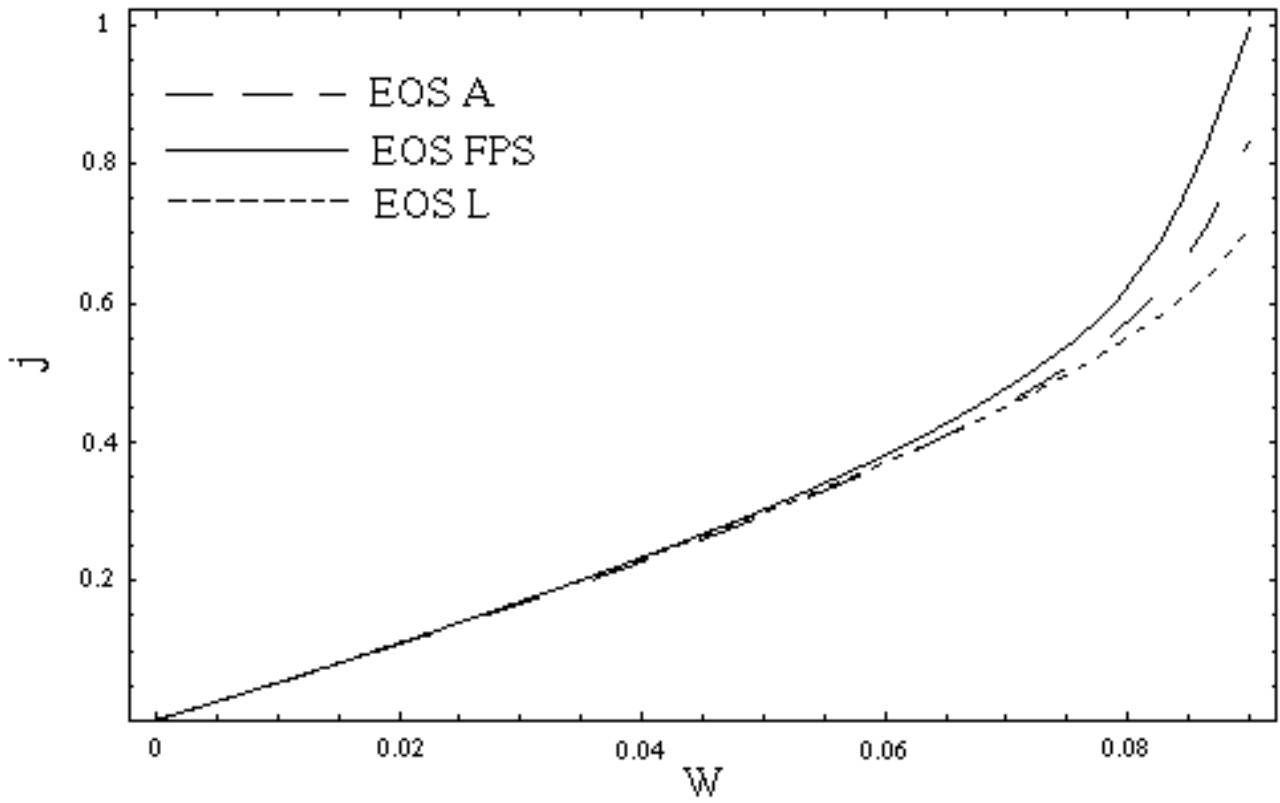
Fig. 6. The function $j =j(W)$ for MMNS
with EOS A, EOS FPS, and EOS L (dashed, solid,
dotted curves respectively).

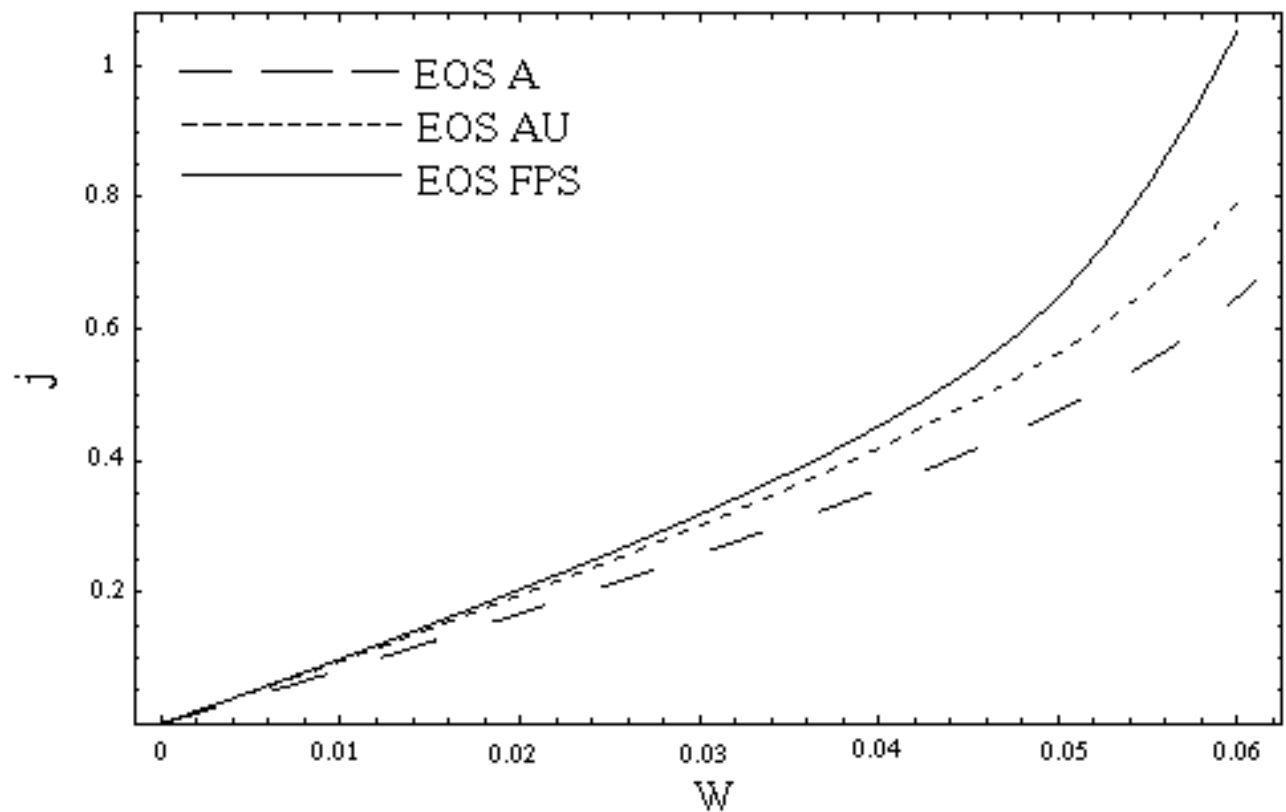
Fig. 7. The function $j =j(W)$ for 1.4   $M_{\odot}$  normal sequences in the 
static limit and with EOS A, EOS AU  and EOS FPS (dashed, dotted, solid curves  
respectively).

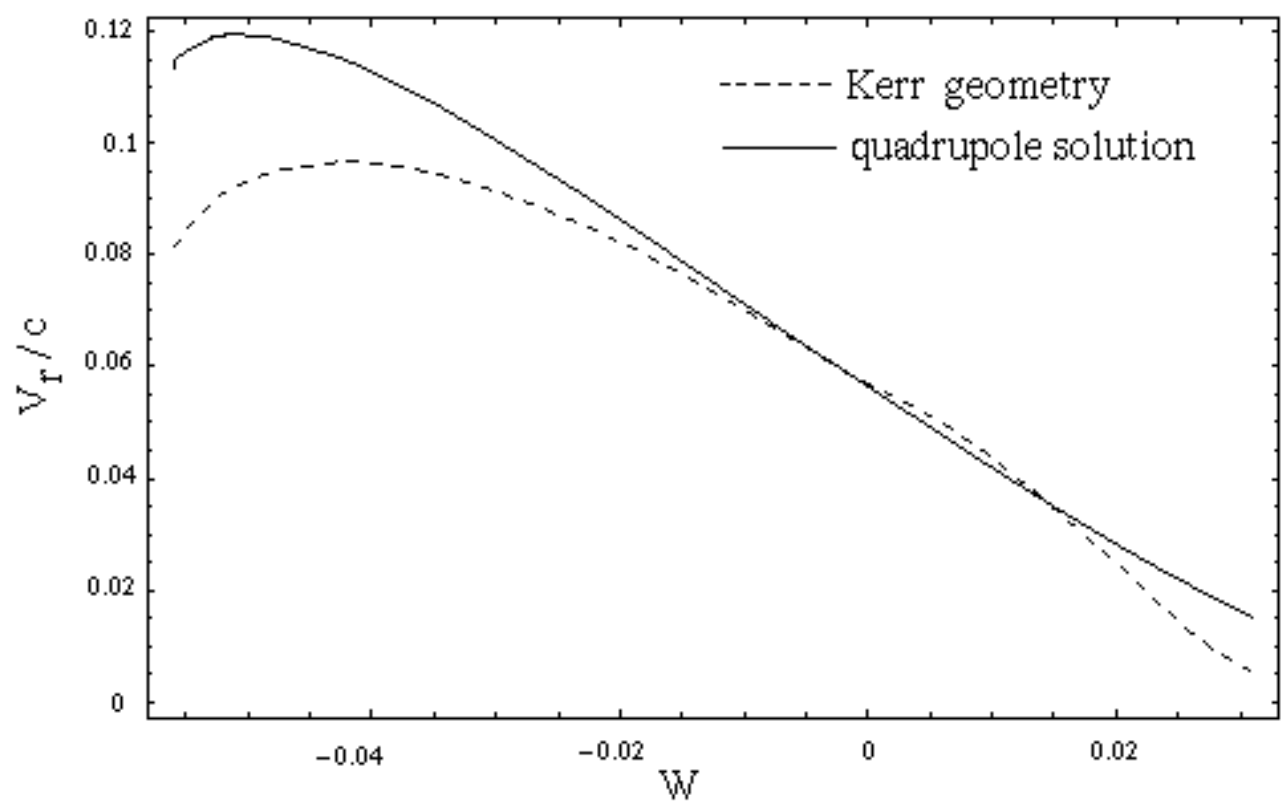
Fig. 8. The radial physical velocity component of the particles (in the 
ZAMO tetrad) that fall from the  marginally stable orbit on the stellar surface 
versus the dimensionless angular velocity $W$ for a 1.4  $M_{\odot}$ normal 
sequence with EOS A in a realistic gravitational field (solid curve
). Dotted
curve  corresponds to the Kerr solution.

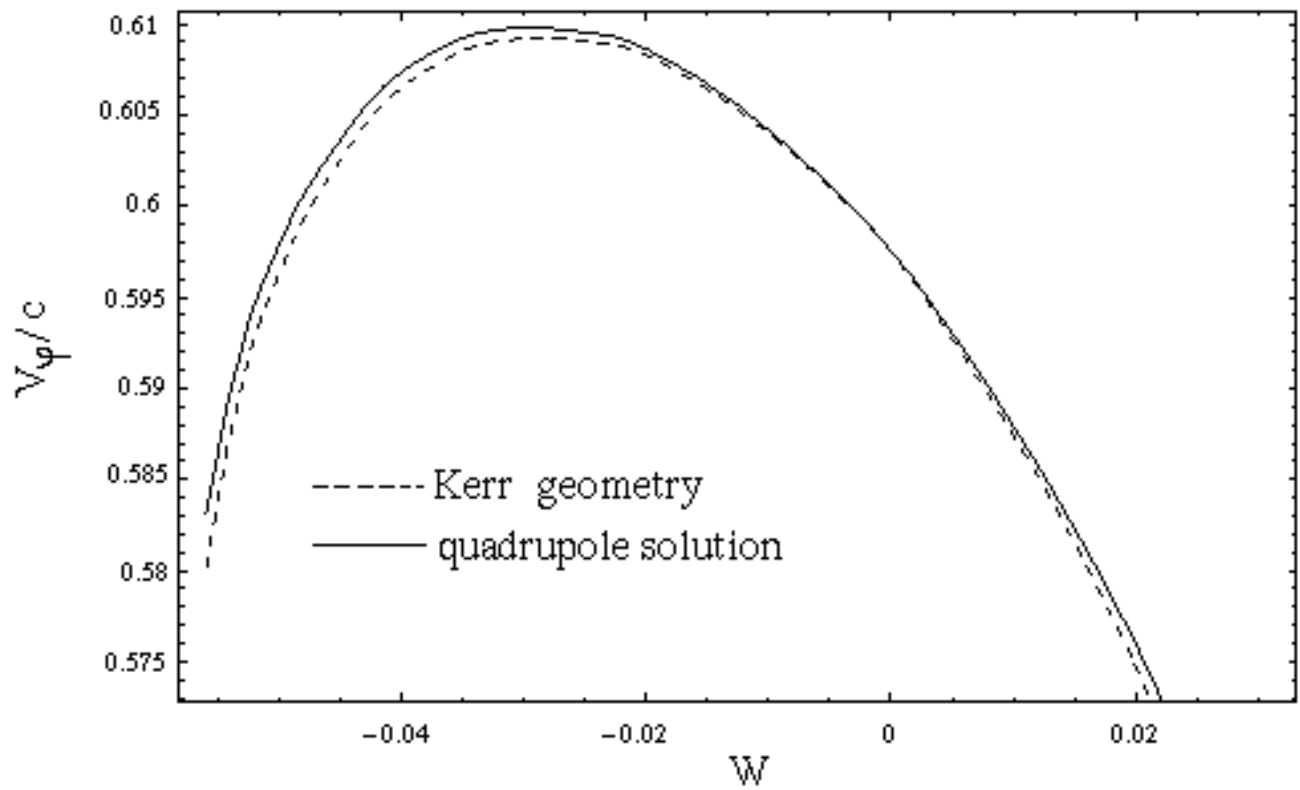
Fig. 9. The same as Fig. 8 for the azimuthal physical velocity 
component.

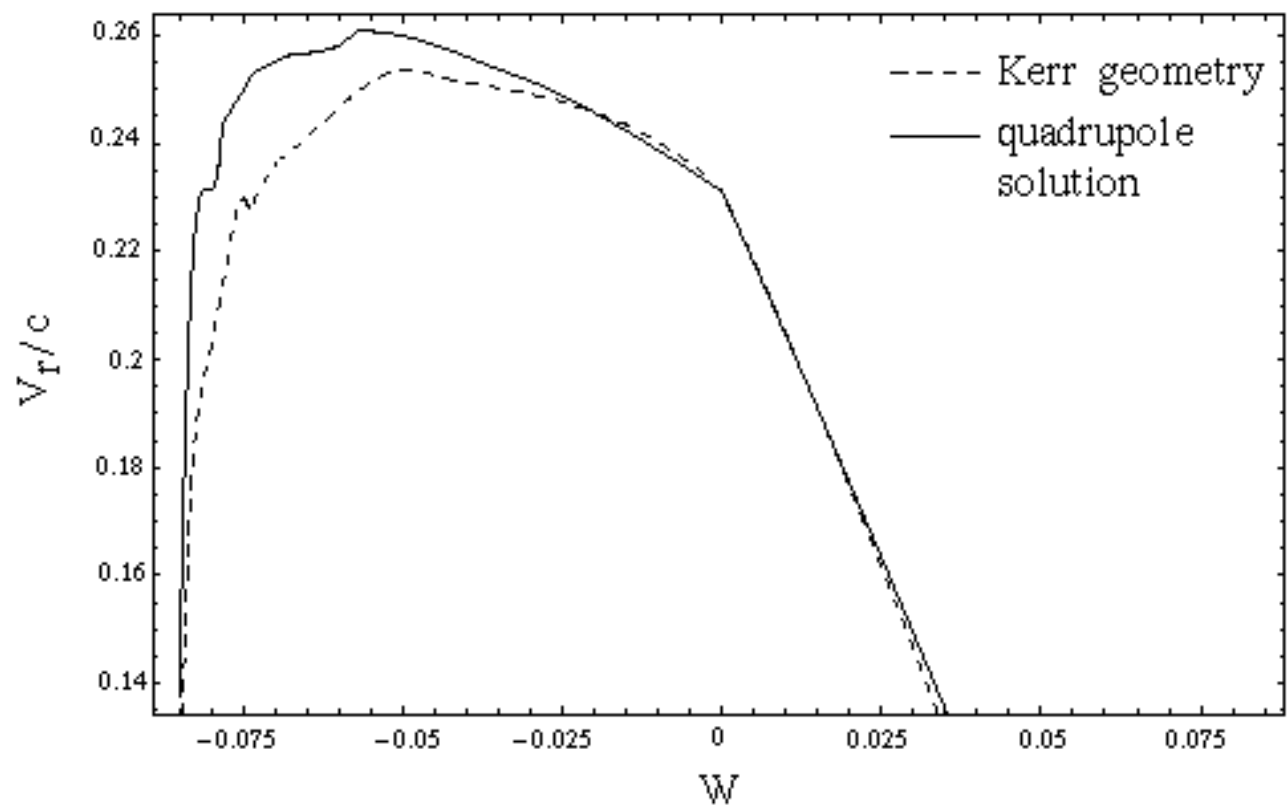
Fig. 10. The radial physical velocity component of the particles (in the 
ZAMO tetrad) that fall from the  marginally stable orbit on the stellar surface 
versus the dimensionless angular velocity $W$ for a MMNS with EOS A in
a realistic gravitational field (solid curve ). Dotted curve 
 corresponds to the Kerr solution.

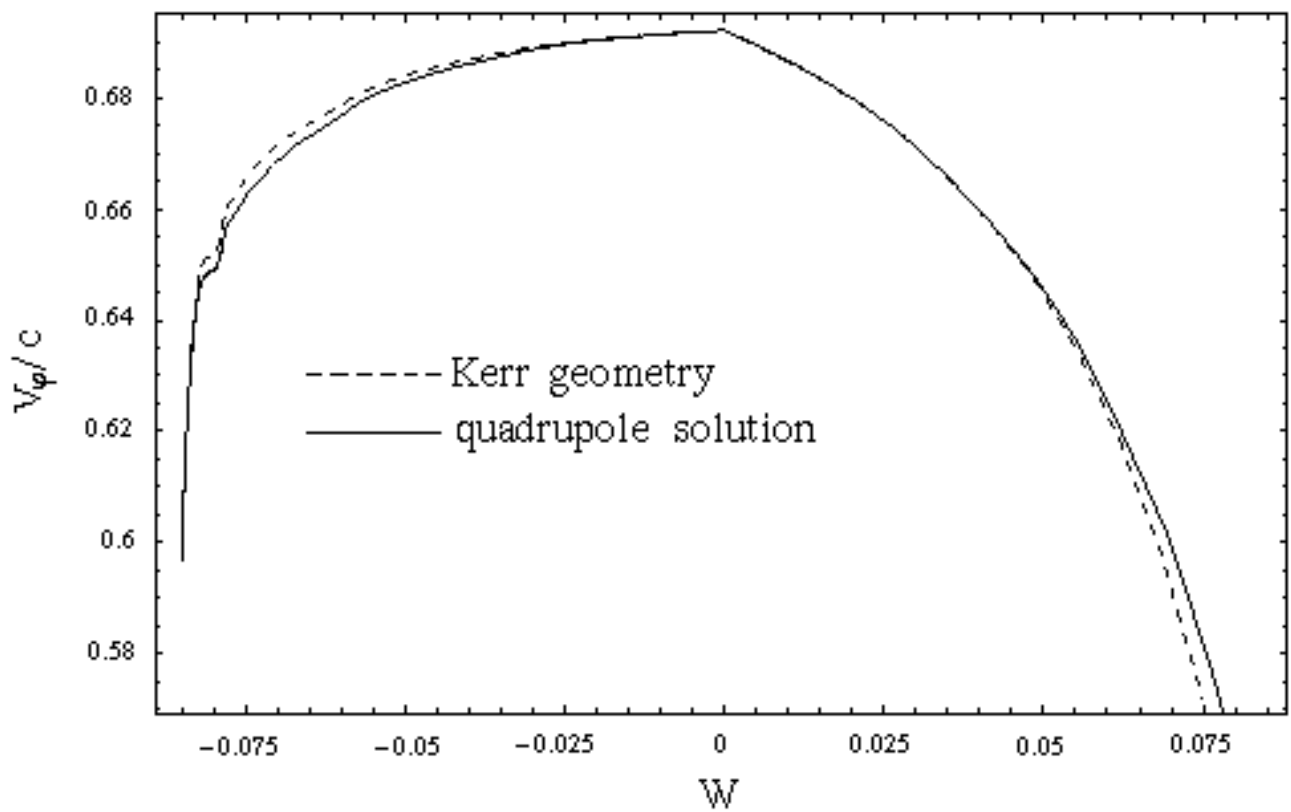
Fig. 11. The same as Fig. 10 for the azimuthal physical velocity 
component.

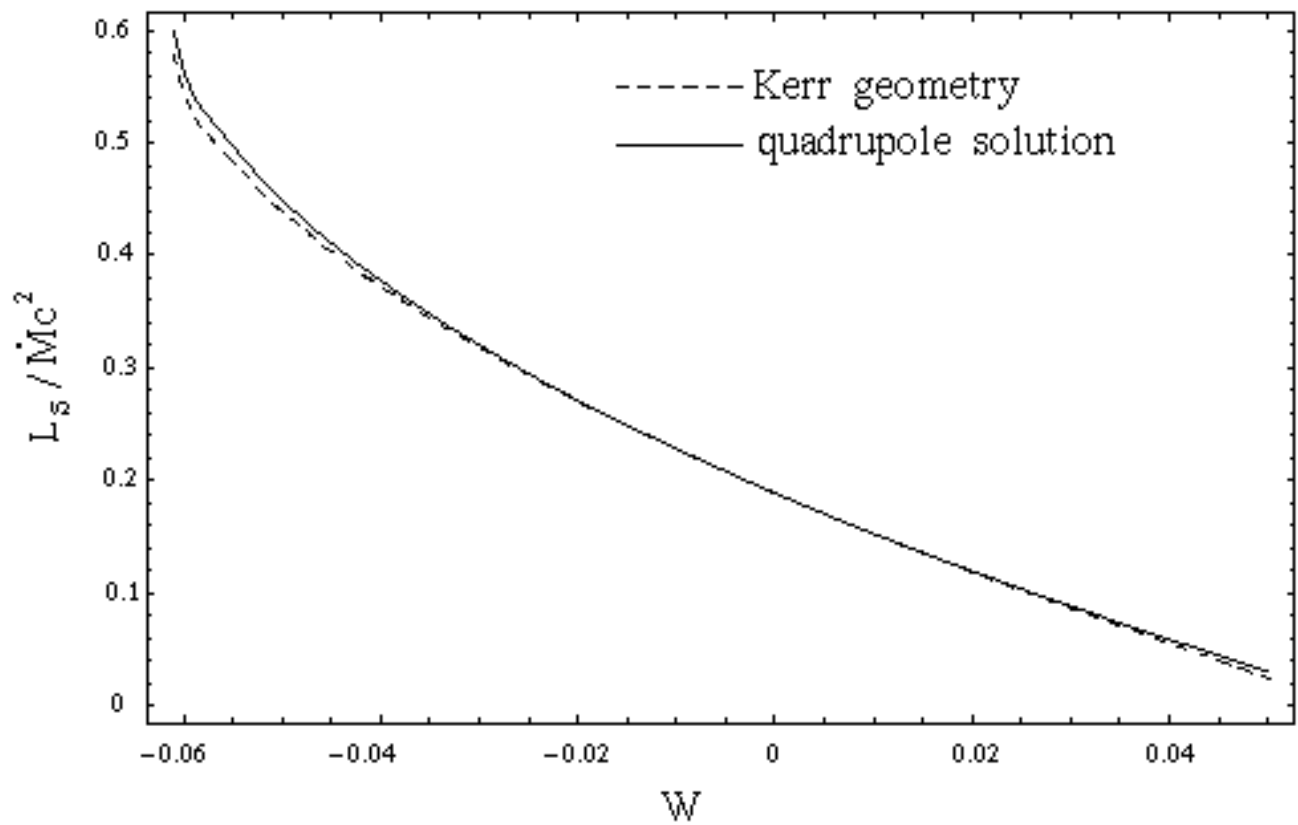
Fig. 12. The luminosity $L_s$ from the accretion boundary layer versus 
the dimensionless angular velocity $W$ for a 1.4  $M_{\odot}$  normal sequence with 
EOS A against the background of a realistic gravitational field for $R < 
R_*$ (solid curve ). Dotted curve  corresponds to the Kerr solution.

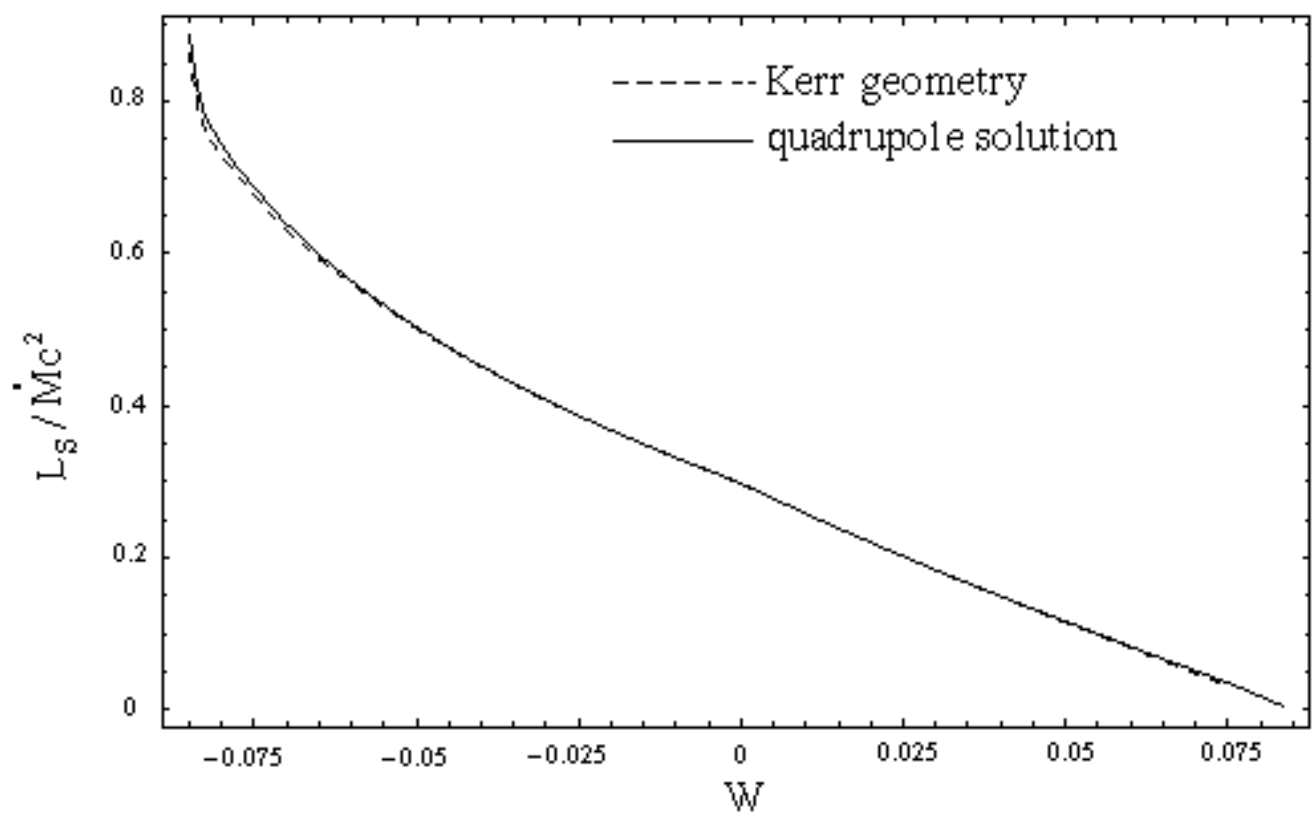
Fig. 13. The luminosity $L_s$ from the accretion boundary layer versus 
the dimensionless angular velocity $W$ for a MMNS 
with EOS A against the background of a realistic gravitational field for 
$R < R_*$ (solid curve ).  Dotted curve  corresponds to the Kerr solution

Fig. 14 The ratio of the luminosities from the accretion boundary 
layer and the entire disc versus 
the dimensionless angular velocity $W$ for a 1.4  $M_{\odot}$  normal sequence with 
EOS A against the background of a realistic gravitational field for $R < 
R_*$ (solid curve ).  Dotted curve  corresponds to the Kerr solution.

Fig. 15 The ratio of the luminosities from the accretion boundary 
layer and the entire disc versus 
the dimensionless angular velocity $W$ for a MMNS with 
EOS A against the background of a realistic gravitational field for $R < 
R_*$ (solid curve ). Dotted curve  corresponds to the Kerr solution. 
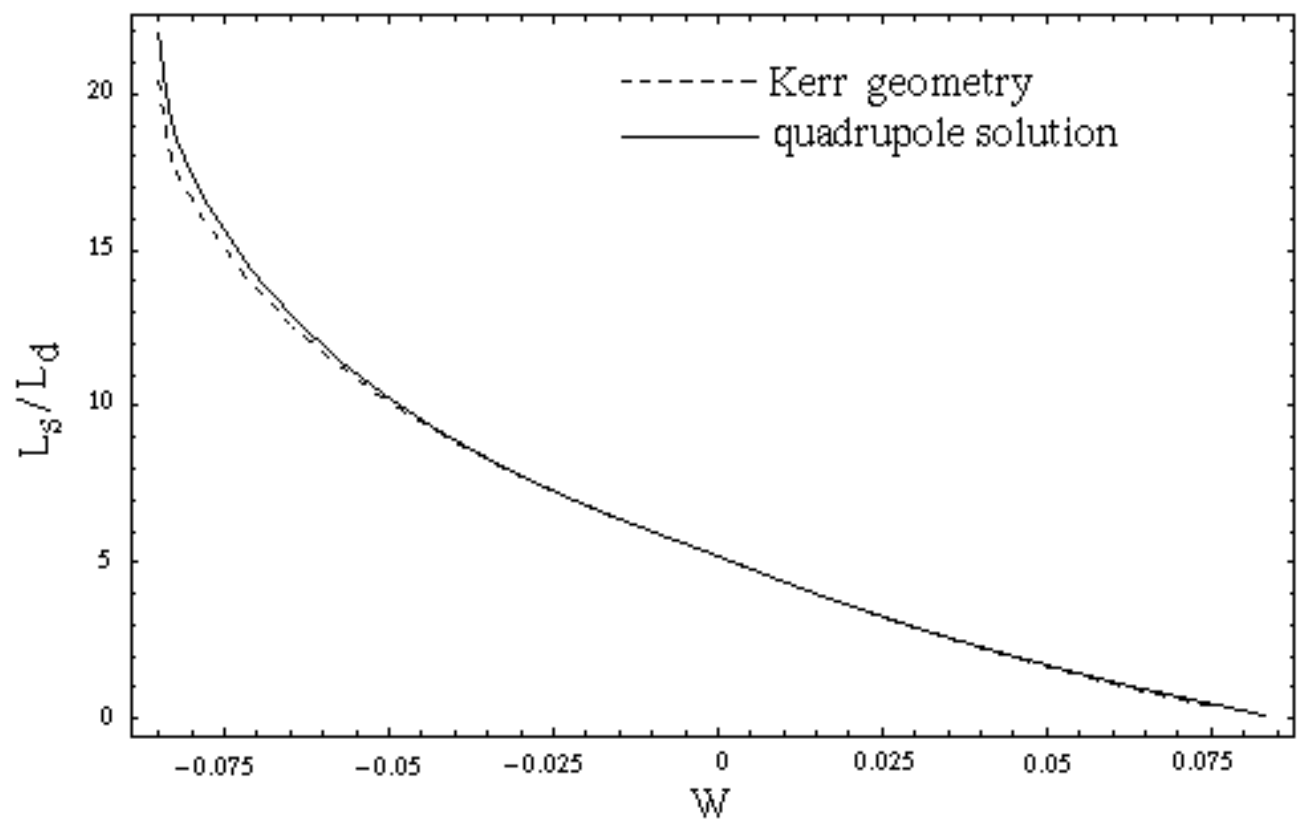
\end{document}